\documentclass[%
reprint,
superscriptaddress,
showpacs,preprintnumbers,
 amsmath,amssymb,
 aps,
 longbibliography,
prb,
floatfix,
]{revtex4-2}

\usepackage{xspace}
\usepackage{graphicx}% Include figure files
	\graphicspath{{./figures/}}
\usepackage{dcolumn}% Align table columns on decimal point
\usepackage{bm}% bold math
\usepackage{color}
\usepackage{hyperref}
\hypersetup{
    unicode=false,          % non-Latin characters in Acrobat’s bookmarks
    pdftoolbar=true,        % show Acrobat’s toolbar?
    pdfmenubar=true,        % show Acrobat’s menu?
    pdffitwindow=false,     % window fit to page when opened
    pdfstartview={FitH},    % fits the width of the page to the window
    pdftitle={Phonon anomalies in FeS},    % title
    pdfnewwindow=true,      % links in new window
    colorlinks=true,       % false: boxed links; true: colored links
    linkcolor=blue,          % color of internal links (change box color with linkbordercolor)
    citecolor=blue,        % color of links to bibliography
    filecolor=blue,      % color of file links
    urlcolor=blue,       % color of external links
    plainpages=false,        % hat Einfluss auf die Seitennummerierung
}

\newcommand{\etal}{\textit{et al.}\xspace}

\newcommand{\Ts}{\ensuremath{T_\mathrm{s}}\xspace}
\newcommand{\Tc}{\ensuremath{T_\mathrm{c}}\xspace}

\newcommand{\Tfm}{\ensuremath{T_\mathrm{{fluct,max}}}\xspace}
\newcommand{\BFCA}{\ensuremath{\mathrm{Ba(Fe}_{1-x}\mathrm{Co}_{x})_{2}\mathrm{As}_{2}}\xspace}
\newcommand{\FESES}{\ensuremath{\mathrm{FeSe}_{1-x}\mathrm{S}_{x}}\xspace}

\newcommand{\Alg}{\texorpdfstring{\ensuremath{A_{1g}}\xspace}{A1g}}
\newcommand{\Algph}{\texorpdfstring{\ensuremath{A_{1g}^{ph}}\xspace}{A1g}}

\newcommand{\AZg}{\texorpdfstring{\ensuremath{A_{2g}}\xspace}{A2g}}
\newcommand{\Blg}{\texorpdfstring{\ensuremath{B_{1g}}\xspace}{B1g}}
\newcommand{\Blgph}{\texorpdfstring{\ensuremath{B_{1g}^{ph}}\xspace}{B1g}}
\newcommand{\BZg}{\texorpdfstring{\ensuremath{B_{2g}}\xspace}{B2g}}

\newcommand{\grd}{$^{\circ}$\xspace}

\newcommand{\wn}{\ensuremath{\rm cm^{-1}}\xspace}
\begin{document}

%%%%%%%%%%%%%%%%%%%%%%%%%%%%%%%%%%%%
%\begin{CJK*}{GBK}{}%%%%%%%%%%%%%%%%%
%%%%%%%%%%%%%%%%%%%%%%%%%%%%%%%%%%%%

%%% authors %%%
%%% WMI Raman: A. Baum, L. Peis, R. Stumberger, R. Hackl
%%% WMI ctd.: R. Hosseinian Ahangharnejad (FeSe only), M. Mitschek (FeS only) -> maybe acknowledgements?
%%% IPB Raman: N. Lazarevic, A. Milosavljevic, Z. Popovic
%%% IPB FeS ellipsometry: M. Scepanovic, M. Grujic-Brojcin -> maybe acknowledgements?
%%% Samples Fe(Se)S: A. Wang, C. Petrovic,
%%% Samples: FeSe measured WMI: T. Wolf (KIT), P. Adelmann (KIT)

\title{Evolution of lattice, spin, and charge properties across the phase diagram of \FESES}
\date{\today}

\author{N.~Lazarevi\'{c}}
\altaffiliation{contributed equally}
\affiliation{Center for Solid State Physics and New Materials, Institute of Physics Belgrade, University of Belgrade, Pregrevica 118, 11080 Belgrade, Serbia}

\author{A.~Baum}
\altaffiliation{contributed equally}
\affiliation{Walther Meissner Institut, Bayerische Akademie der Wissenschaften, 85748 Garching, Germany}
\affiliation{Fakult\"at f\"ur Physik, Technische Universit\"at M\"unchen, 85478 Garching, Germany}
\author{A.~Milosavljevi\'{c}}
\affiliation{Center for Solid State Physics and New Materials, Institute of Physics Belgrade, University of Belgrade, Pregrevica 118, 11080 Belgrade, Serbia}

\author{L.~Peis}
%\altaffiliation{contributed equally}
%\altaffiliation{Fakult\"at f\"ur Physik, Technische Universit\"at M\"unchen, 85478 Garching, Germany}
\affiliation{Walther Meissner Institut, Bayerische Akademie der Wissenschaften, 85748 Garching, Germany}
\affiliation{Fakult\"at f\"ur Physik, Technische Universit\"at M\"unchen, 85478 Garching, Germany}

\author{R. Stumberger}
%\altaffiliation{contributed equally}
%\altaffiliation{Fakult\"at f\"ur Physik, Technische Universit\"at M\"unchen, 85478 Garching, Germany}
\affiliation{Walther Meissner Institut, Bayerische Akademie der Wissenschaften, 85748 Garching, Germany}
\affiliation{Fakult\"at f\"ur Physik, Technische Universit\"at M\"unchen, 85478 Garching, Germany}
\author{J. Bekaert}
\affiliation{Department of Physics, University of Antwerp, Groenenborgerlaan 171, B-2020 Antwerp, Belgium}
\author{A. \v{S}olaji\'{c}}
\affiliation{Center for Solid State Physics and New Materials, Institute of Physics Belgrade, University of Belgrade, Pregrevica 118, 11080 Belgrade, Serbia}
\author{J. Pe\v{s}i\'{c}}
\affiliation{Center for Solid State Physics and New Materials, Institute of Physics Belgrade, University of Belgrade, Pregrevica 118, 11080 Belgrade, Serbia}
\author{Aifeng Wang}
\affiliation{School of Physics, Chongqing University, Chongqing 400044, China}

\author{M. \v{S}\'{c}epanovi\'{c}}
\affiliation{Center for Solid State Physics and New Materials, Institute of Physics Belgrade, University of Belgrade, Pregrevica 118, 11080 Belgrade, Serbia}
\author{M. V. Milo\v{s}evi\'{c}}
\affiliation{Department of Physics, University of Antwerp, Groenenborgerlaan 171, B-2020 Antwerp, Belgium}
\author{C.~Petrovic}
\affiliation{Condensed Matter Physics and Materials Science Department, Brookhaven National Laboratory, Upton, NY 11973-5000, USA}
\author{Z.V.~Popovi\'{c}}
\affiliation{Center for Solid State Physics and New Materials, Institute of Physics Belgrade, University of Belgrade, Pregrevica 118, 11080 Belgrade, Serbia}
\affiliation{Serbian Academy of Sciences and Arts, Kneza Mihaila 35, 11000 Belgrade, Serbia}
\author{R.~Hackl}
\affiliation{Walther Meissner Institut, Bayerische Akademie der Wissenschaften, 85748 Garching, Germany}
\affiliation{Fakult\"at f\"ur Physik, Technische Universit\"at M\"unchen, 85478 Garching, Germany}
\affiliation{IFW Dresden, Helmholtzstr. 20, 01069 Dresden, Germany}

%%%%%%%%%%%%%%%%%%%%%%%%%%%%%%%%%%%%%%%%%%%%%%%%%%%%%%%%%%%%%%%%%%%%%%%%%%%%
\begin{abstract}
A Raman scattering study covering the entire substitution range of the  \FESES solid solution is presented. Data were taken as a function of sulfur concentration $x$ for $0\le x \le 1$, of temperature and of scattering symmetry. All type of excitations including phonons, spins and charges are analyzed in detail. It is observed that the energy and width of iron-related \Blg phonon mode vary continuously across the entire range of sulfur substitution. The \Alg chalcogenide mode disappears above $x=0.23$ and reappears at a much higher energy for $x=0.69$. In a similar way the spectral features appearing at finite doping in \Alg symmetry vary discontinuously. The magnetic excitation centered at approximately 500\,\wn disappears above $x=0.23$ where the \Alg lattice excitations exhibit a discontinuous change in energy. The low-energy mode associated with fluctuations displays maximal intensity at the nemato-structural transition and thus tracks the phase boundary. 
\end{abstract}
%%%%%%%%%%%%%%%%%%%%%%%%%%%%%%%%%%%%%%%%%%%%%%%%%%%%%%%%%%%%%%%%%%%%%%%%%%%%%%%%%%%%%%%%%%%%%%%
\pacs{%
%78.30.-j, %Infrared and Raman spectra
%74.72.-h, %cuprate superconductors
74.70.Xa, %pnictides and chalcogenides
%75.10.Jm, %quantized spin models including frustration
%74.20.Mn, %nonconventional mechanisms
74.25.nd %Raman and optical spectroscopy (of superconductors)
%63.20.K-, %Phonon interactions
75.25.Dk	%Orbital, charge, and other orders, including coupling of these orders
}
\maketitle

%%%%%%%%%%%%%%%%%%%%%%%%%
%\end{CJK*}%%%%%%%%%%%%%%
%%%%%%%%%%%%%%%%%%%%%%%%%

%%%%%%%%%%%%%%%%%%%%%%%%%%%%%%%%%%%%%%%%%%%%%%%%%%%%%%%%%%%%%%%%%%%%%%%%%%%%%%%%%%%%%%%%%%%%

\section{Introduction}

Iron-based compounds are widely believed to host unconventional superconductivity, thus being similar to cuprates or heavy Fermion systems. All are characterized by competing phases including magnetism, crystal symmetry breaking or nematicity and fluctuations of charge and spin prior to superconductivity \cite{Scalapino:2012,Fradkin:2015,Lederer:2015}. While long range magnetic ordering was found in the majority of the compounds, it is absent in the binary compound FeSe. Yet a nematic and structural phase transition occurs simultaneously at 90\,K \cite{Boehmer2015_PRL114_027001, McQueen2009_PRL103_057002, Watson2015_PRB91_155106}. Below $\Tc=9$\,K superconductivity is observed \cite{Hsu2008_PNAS105_14262}. Upon applied pressure \Tc increases to approximately 37\,K  \cite{Medvedev2009_NM8_630}. By substituting sulfur for selenium, the transition temperature to the nematic phase is suppressed to zero for $x \sim 0.2$ \cite{Sato1227}, suggesting the existence of a quantum critical point (QCP), and a depression of \Tc to approximately 2\,K. For $x > 0.2$, \Tc increases again and reaches  5\,K at $x=1$ \cite{Lai2015_JACS137_10148}. Surprisingly enough, FeS displays a metallic variation of the resisitivity and a high residual resistivity ratio RRR of approximately 30, and  neither structural nor nematic phase transitions occur \cite{Pachmayr2016_CC52_194}. Thus, \FESES uniquely offers access to instabilities and critical points and the disappearance thereof while superconductivity survives.

FeSe and FeS are iso-structural, thus providing us with the opportunity to probe the evolution of competing order by iso-electronic substitution. We wish to address the question as to which extent the properties and specifically superconductivity are interrelated with the other instabilities and how the electronic properties affect the phonons. We employ inelastic light scattering to probe evolution with composition of lattice spin and charge excitations in \FESES \cite{Lazarevic_2020}. We identify the \Alg and \Blg modes, a two-phonon scattering process as well as additional modes that can be traced to either defect-induced or second-order scattering. The obtained experimental results are in good agreement with numerical calculations. Phonons self-energy temperature dependence supports the results reported in Refs.~\onlinecite{Holenstein2016_PRB93_140506,Kirschner2016_PRB94_134509} where emerging short range magnetic order at approximately $20$\,K was reported.

\section{Experiment}
\label{sec:exp}

Single crystals of \FESES were synthesized as described elsewhere.\cite{Wang2016_PRB94_094506} Up to $x=0.22$ the samples were prepared by vapor transport. For $x>0.22$ only the hydrothermal method yields homogeneous single crystals. Before the experiment the samples were cleaved in air.

Inelastic light scattering on phonons was performed using a Tri Vista 557 Raman spectrometer with the first two monochromators coupled subtractively and the grating combination 1800/1800/2400 grooves/mm. For excitation a Coherent Verdi G solid state laser was used emitting at 532\,nm. The samples were mounted in a KONTI CryoVac continuous helium flow cryostat having a 0.5\,mm thick window. The vacuum was pumped to the range of $10^{-6}$\,mbar using a turbo molecular pump. The laser was focused to a spot size of approximately $8\,\mu{\rm m}$ using a microscope objective lens with $\times$50 magnification. In this back-scattering configuration the plane of incidence coincides with crystallographic $c$-axis. All Raman spectra were corrected for the Bose factor.

Fluctuations and two magnon excitations were probed with a calibrated scanning spectrometer. The samples were attached to the cold finger of a He-flow cryostat having a vacuum of better than $10^{-6}$\,mbar. A diode-pumped solid state laser emitting  at 575\,nm (Coherent GENESIS) was used as an excitation source. The laser beam was focused on the sample at an angle of incidence of 66$^\circ$. Polarization and power of the incoming light were adjusted in a way that the light inside the sample had the proper polarization state and a power of $P_a=4$\,mW independent of polarization. All four symmetries of the D\textsubscript{4h} group,  \Alg, \AZg, \Blg, and \BZg,  can be accessed using appropriate in-plane polarizations of the incident and scattered light.

\begin{figure}
  \centering
  \includegraphics[width=85mm]{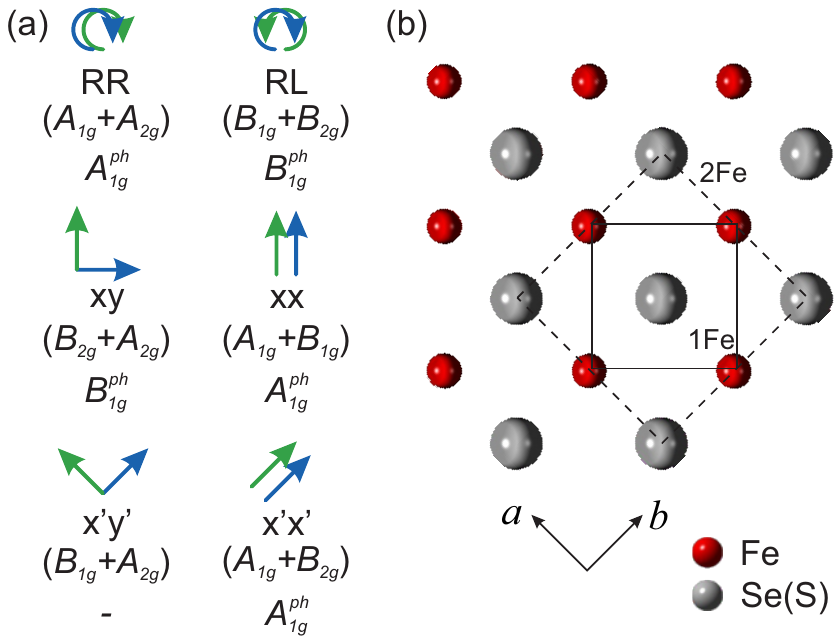}
  \caption{Crystal structure and selection rules for FeSe(S). Solid and dashed lines represent the 1\,Fe and the crystallographic 2\,Fe unit cell, respectively. The crystallographic axes are $a$ and $b$. In FeSe and FeS only one \Alg and one \Blg phonon is expected as indicated by \Algph and \Blgph, respectively. The symmetries projected with the polarizations indicated symbolically with respect to the 1\,Fe cell are relevant for electronic and spin excitations. The symmetries of the phonons are in brackets.}
  \label{fig:Selectionrules}
\end{figure}

The selection rules are dictated by the crystal structure. Here, only polarizations in the  $ab$ plain are relevant, as shown in in Fig.~\ref{fig:Selectionrules}, with solid and dashed lines representing 1\,Fe and 2\,Fe unit cells, respectively. For the tetragonal system there are six principal scattering geometries, each of which probing two symmetry channels. We align our laboratory system with the 1\,Fe unit cell. As a consequence, the \Blg phonon (\Blgph) is observable in $xy$ configuration which corresponds to the \BZg symmetry channel in the 2\,Fe cell [Fig.~\ref{fig:Selectionrules}]. We decided to use this orientation since our main focus here are electronic and spin excitations for which the 1\,Fe unit cell is more appropriate. \Algph is the fully symmetric in-phase Se(S) mode with elongations along the $c$-axis, \Blgph corresponds to the out-of-phase vibration of the Fe atoms parallel to the $c$-axis.

 %%%%%%%%%%%%%%%%%%%%%%%%%%%%%%%%%%%%%%%%%%%%%
%%%%%%%%%%%%%%%%%%%%%%%%%%%%%%%%%%%%%%%%%%%%%

%Calibrated customized Raman scattering equipment was used for the experiment. The samples were attached to the cold finger of a He-flow cryostat having a vacuum of approximately $5\cdot10^{-5}$\,Pa. For excitation we used a diode-pumped solid state laser emitting at 575\,nm (Coherent GENESIS).
%Polarization and power of the incoming light were adjusted in a way that the light inside the sample had the proper polarization state and, respectively, a power of typically $P_a=4$\,mW independent of polarization. The crystallographic axes are $a$ and $b$ with $|a| = |b|$. The $c$-axis is parallel to the optical axis. $a^{\prime}$ and $b^{\prime}$ are rotated by 45\grd w.r.t. $a$ and $b$. The laser beam reached the sample at an angle of incidence of 66$^\circ$ and was focussed to a spot of approximately $50\,\mu{\rm m}$ diameter.  By choosing proper in-plane polarizations of the incident and scattered light the four symmetry channels \Alg, \AZg, \Blg, and \BZg of the D\textsubscript{4h} space group can be accessed.
%Additionally, for the large angle of incidence, exciting photons being polarized along the $b$-axis have a  finite $c$-axis projection, and the \Eg symmetry can also be accessed. For the symmetry assignment we use the 2\,Fe unit cell (crystallographic unit cell).

\section{Results and Discussion}
\label{sec:results}
\subsection{Lattice excitations}

\begin{figure}
  \centering
  \includegraphics[width=85mm]{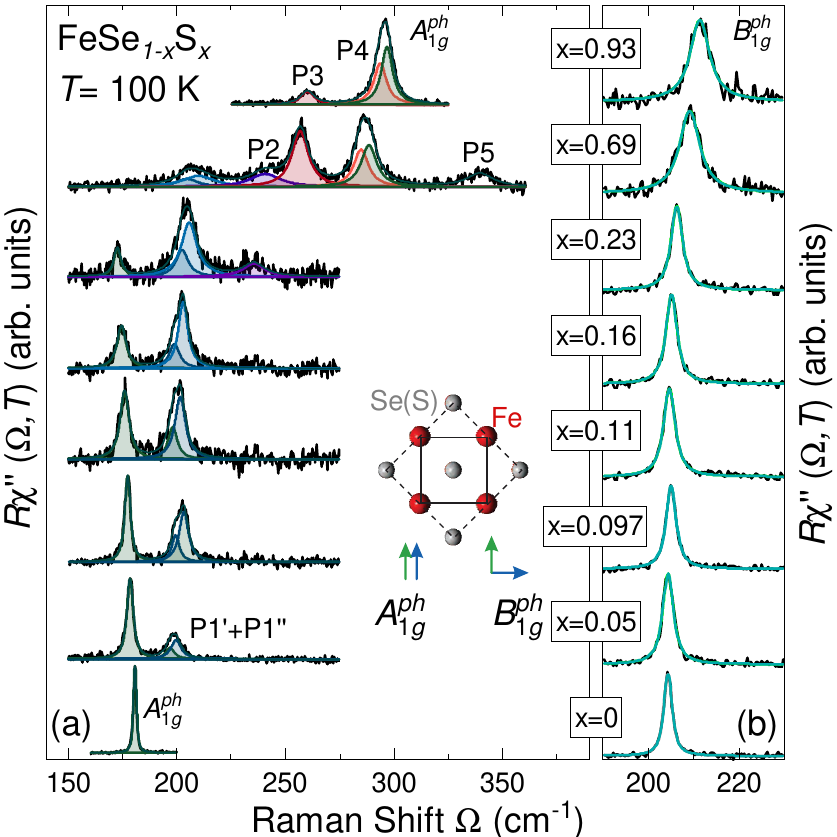}
  \caption{Phonon spectra of \FESES measured at 100\,K. We show $xx$ and $xy$ spectra where $x$ and $y$ are rotated by 45\grd with respect to 2\,Fe unit cell, as indicated in the inset, and project  \Algph and \Blgph, respectively. (a) \Algph spectra. Only for pure FeSe ($x=0$) a single line is observed at the \Alg energy of 165\,\wn expected from lattice dynamics. Above $x=0.23$ the Se(S) vibration becomes unobservable and reappears only for $x\ge0.69$ at a much higher energy of approximately 290\,\wn similar to that in pure FeS. The peaks other than Raman-active phonons are labeled P1--P5 with increasing energy. Solid lines represent the best fits to the data using Voigt profiles. (b) \Blgph spectra. Energy and linewidth vary continuously with sulphur content.}
  \label{fig:Phonons}
\end{figure}

\begin{figure}
  \centering
  \includegraphics[width=85mm]{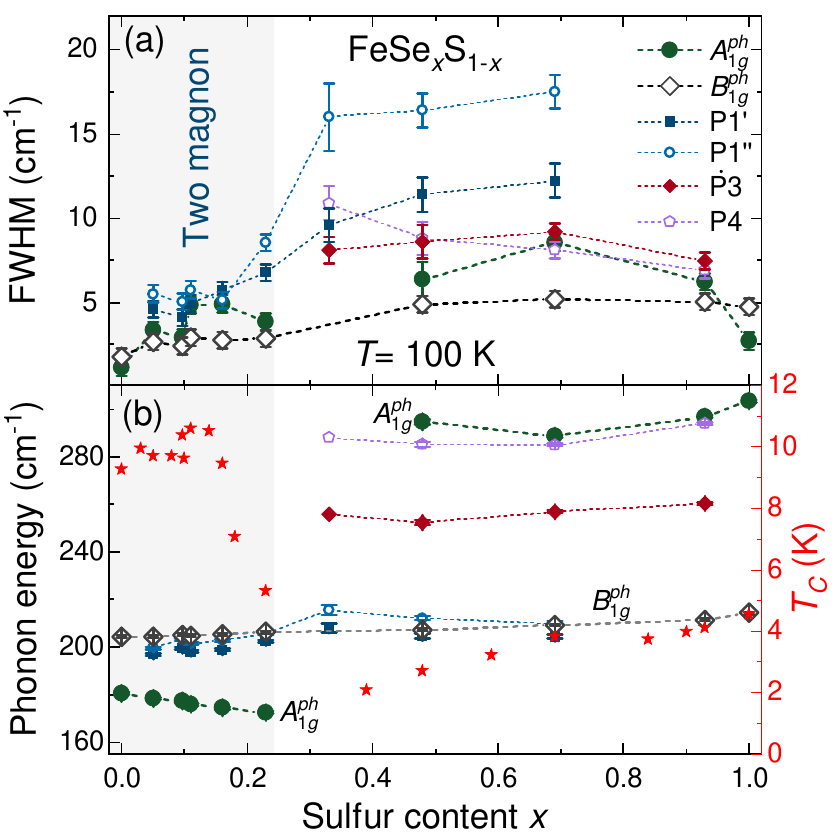}
  \caption{Energies and linewidths of the Raman active modes and \Tc in \FESES as a function of sulfur content $x$ at 100\,K. (a) Peak widths (FWHM) and (b) energies as obtained from the fits (left axis). The \Tc values of the corresponding solid solution are taken from Ref.  \cite{WangA:2020}.}
  \label{fig:Energy_linewidth}
\end{figure}

First, the focus is placed on lattice excitations observable in the $xx$ and $xy$ scattering configuration projecting \Algph and \Blgph in the spectral range characteristic for phonons. Fig.~\ref{fig:Phonons} shows the evolution of the spectra with doping $x$, where $x$ indicates the sulfur concentration. Additional spectra for $x=0.48$ and, for convenience, $x=1$ are shown in Fig. \ref{Afig:more}. In order to minimize the thermal broadening of the modes while staying above the nematic phase transition, the spectra were recorded at 100\,K. In pure stoichiometric compounds only one \Algph and one \Blgph phonon mode is expected (see Fig.~\ref{fig:Selectionrules}). This is indeed the case for FeSe [Fig.~\ref{fig:Phonons}(a) bottom] as described by Gnezdilov \etal \cite{Gnezdilov2013_PRB87_144508} and corroborated here. In contrast, in pure FeS ($x=1$) additional modes exist in the $xx$ spectrum which were assigned to two-phonon scattering (265\,\wn) and a projection of the phonon density of states (PDOS) ($\sim$300\,\wn) \cite{Baum:2018} as reproduced in Fig. \ref{Afig:more}~(b). The $xy$ spectra show only the \Blgph mode for all doping levels displayed here (see also Fig. \ref{Afig:more}(a)). It hardens monotonously and exhibits a weak maximum of the linewidth at $x=0.69$ and $x=0.93$ highlighting the effect of disorder as summarized in Fig.~\ref{fig:Energy_linewidth} (a) and (b). 

The $xx$ spectra display a much more complex doping dependence. Upon substituting only a small amount of sulfur ($x=0.05$) for selenium an additional structure appears at about 200\,\wn [Fig.~\ref{fig:Phonons}~(a)]. Closer inspection of the FeSe$_{0.95}$S$_{0.05}$ spectra reveals that this feature consists of two peaks denoted as P$1^\prime$ and P$1^{\prime\prime}$. With increasing $x$, these structures gain intensity and harden slightly, whereas the \Alg phonon softens, gradually loses intensity, and becomes undetectable at concentrations above $x=0.23$. It reappears as a clear peak only for $x\geq 0.69$ at a much higher energy characteristic for FeS \cite{Baum:2018} and possibly as a remnant structure in the spectrum for $x=0.48$ [Fig. \ref{Afig:more}~(a)]. As in FeS the \Algph peak overlaps with a weaker structure which is compatible with the PDOS (P4). At $x=0.69$ P4 is approximately as strong as the \Alg phonon. Here (and at $x=0.48$, Fig. \ref{Afig:more}(a)) there is also a broad feature at 340\,\wn (P5). For $x=0.93$ similar to $x=1$ there is another structure at 250\,\wn (P3) which gains intensity toward $x=0.69$ where it has a weak companion at 235\,\wn (P2) being present down to $x=0.23$. As expected, the increase of crystalline disorder due to substitution leads to a broadening of all observed modes to some maximum value before the trend reverses for compositions close to pure FeS. The widths and energies of the stronger modes are summarized in Fig.~\ref{fig:Energy_linewidth}. As opposed to the \Blgph phonon in $xy$ configuration all modes in $xx$ polarization including the Raman-active phonon depend quasi-discontinuously on substitution.

%%%%%%%%%%%%%%%%%%%%%%%%%%%%%%%%%%%%%%%%%%%%%
%%%%%%%%%%%%%%%%%%%%%%%%%%%%%%%%%%%%%%%%%%%%%

%%%%%%%%%%%%%%%%%%%%%%%%%%%%%%%%%%%%%%%%%%%%%%%%%%%%%%%%%%%%%%
%%%%%%%%%%%%%%%%%%%%%%%%%%%%%%%%%%%%%%%%%%%%%%%%%%%%%%%%%%%%%%%%

This dichotomy of the substitution dependence of the phonon part in $xx$ and $xy$ configuration is the most remarkable effect of this study. Whereas the continuous evolution of the Fe \Blg line by and large tracks the degree of disorder and lattice contraction the Se/S \Alg mode varies counter-intuitively. Naively one would expect a continuous (not necessarily trivial) increase of the phonon frequency and maximal broadening for doping levels around $x=0.5$ similar to what is observed in isotopically substituted semiconductors \cite{Cardona:2005}. However, the line disappears after a continuous loss of intensity at approximately $x=0.23$ and 172\,\wn and reappears (presumably) at $x=0.48$ slightly below 300\,\wn. At low doping the \Alg energy decreases by 4\% although S is lighter than Se by a factor of 2.13 and the lattice contracts. Above $x=0.48$ the energy of the \Alg phonon varies as expected [see Fig. \ref{fig:Energy_linewidth}(b)]. 

The structures appearing in addition to the allowed phonons are rather difficult to interpret in detail. There are essentially two possibilities for intensity to appear in addition to the phonons: defect-induced scattering projecting the PDOS on the site of the defect or overtone (combination) scattering \cite{Turrell:1972}. In FeS one of the peaks (P3) lies in the gap between the acoustic and the optical branches and was therefore assigned to an overtone whereas P4 may originate from the PDOS \cite{Baum:2018}. The two features depend in the same fashion on doping as the \Alg phonon, and the assignment may be maintained. This is plausible on the basis of the PDOS [Fig. \ref{fig_S1}] although the PDOS of a solid solution cannot be calculated straightforwardly. If we argue that the extra lines vary as discontinuously as the phonon, P1$^{\prime}$ and P1$^{\prime\prime}$ would have both an overtone and a PDOS component. Interestingly, P1$^{\prime}$ and P1$^{\prime\prime}$ have the expected doping dependence [see Fig. \ref{fig:Energy_linewidth}(b)]

The anomalous doping dependence of the \Alg phonon may indicate an enhanced electron-phonon coupling which manifests itself also in the linewidth (on top of the inhomogeneous broadening) [Fig. \ref{fig:Energy_linewidth}(a)]. The slightly enhanced electron-phonon coupling may boost \Tc a little bit until the structure becomes unstable and \Tc decreases rapidly for $x>0.16$. There is in fact a kink in the $c/a$ ratio at $x=0.23$ which may be related to the structural instability \cite{WangA:2020}. In a recent preprint the collapse of \Tc is almost precipitous and coincides with the end of the nematic phase \cite{Mitsukami:2021}
and one may speculate about the position of the quantum critical point and its impact. Yet, further work is necessary to finally clarify the issue.

\begin{figure}
  \centering
  \includegraphics[width=85mm]{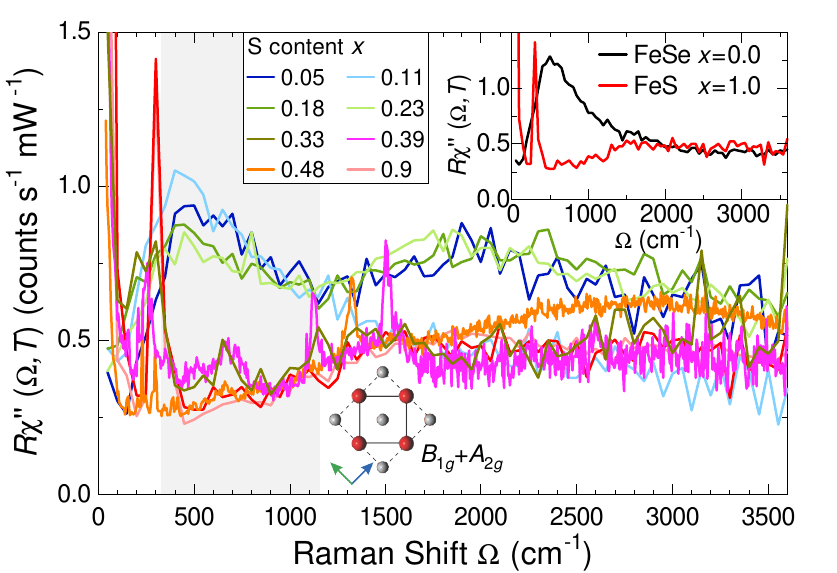}
  \caption{Doping dependence of the high-energy spectra of \FESES in $xy$ (2\,Fe) configuration at 4\,K, except for $x$=0.33, $x$=0.39, $x$=0.48 which were obtained at 100\,K. For the electronic unit cell (full line in Fig. \ref{fig:Selectionrules}) relevant here the \Blg and \AZg symmetries are projected where \AZg is negligibly weak. The doping levels are indicated. The inset compares the high-energy spectra of pure FeSe \cite{Baum:2019} and FeS. The maximum in the range 500\,\wn is compatible with two-magnon scattering \cite{Ruiz:2019} whereas the broad shoulder around 2000\,\wn appearing in three out of ten (including all doping levels) spectra was identified as luminescence by using various laser lines for excitation.}
  \label{fig:magnon}
\end{figure}
%%%%%%%%%%%%%%%%%%%%%%%%%%%%%%%%%%%%%%%%%%%%%%%%%%%%%%%%%%%%%%%%

%%%%%%%%%%%%%%%%%%%%%%%%%%%%%%%%%%%%%%%%%%%
\begin{figure}
  \centering
  \includegraphics[width=85mm]{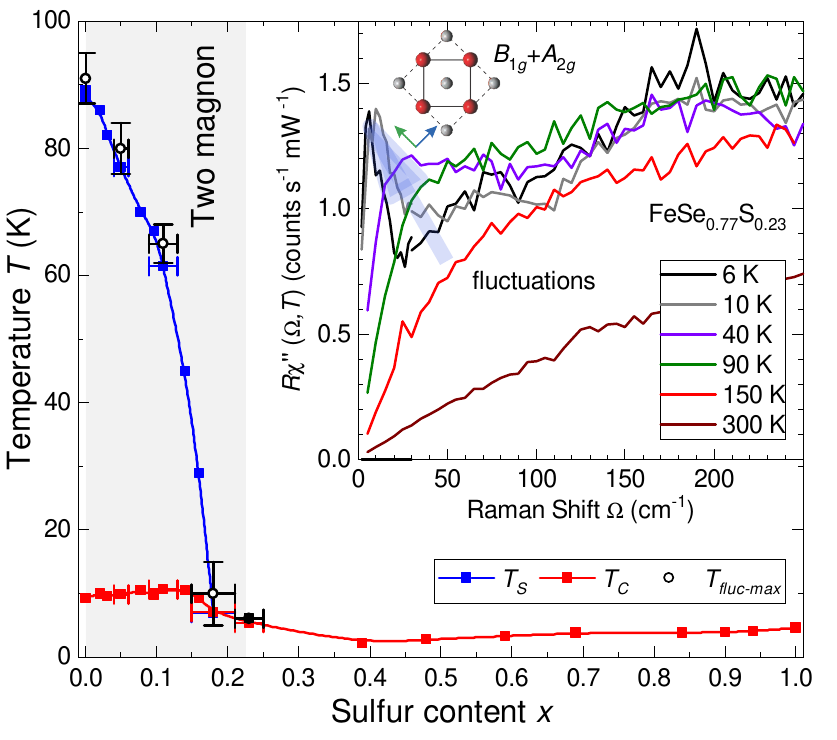}
  \caption{Phase diagram of \FESES  with \Tfm tracing $T_S$ in the region where the two-magnon feature was observed in the Raman scattering experiment. \Tc taken from Ref.  \cite{WangA:2020}. Inset: Low energy Raman spectrum showing fluctuation contribution temperature dependence at $x$=0.23.}
  \label{fig:fluct}
\end{figure}
%%%%%%%%%%%%%%%%%%%%%%%%%%%%%%%%%%%%%%%%%%%%%%%%%%%%%%%%%%%%%%%%

\subsection{Spin excitations and fluctuations}
Second, we focus on the electronic \Blg symmetry channel projected in $x^\prime y^\prime$ (1\,Fe) configuration. Fig.~\ref{fig:magnon} shows the doping dependence of the high energy Raman spectra at approximately 4\,K. The \AZg contribution can be neglected in these materials \cite{Baum:2019}. A broad excitation centered at about 500\,\wn dominates the spectrum at $x=0$ which was interpreted in terms of two magnon scattering \cite{Baum:2019}. Since the ratio of the nearest to the next-nearest-neighbor exchange coupling $J1$ and $J2$ is close to 0.5 \cite{Glasbrenner:2015} the system is a nearly frustrated antiferromagnet. Consequently the two-magnon Raman peak is pushed to energies well below $3\,J1$ \cite{Ruiz:2019}. No comparable feature is observed in FeS (see inset Fig.~\ref{fig:magnon}).  

Upon doping, the two-magnon peak remains relatively robust up to $x=0.23$ and is absent for higher doping levels. This goes in line with the fact that for $x=0$, the Fermi velocity in the $d_{xy}$ band, $v_{\rm F}^{(xy)}$, is significantly smaller than $v_{\rm F}^{(xz)}$ or $v_{\rm F}^{(yz)}$ and increases by only 10\% for $x\le 0.20$. For $x>0.20$ $v_{\rm F}^{(xy)}$ increases significantly towards FeS \cite{Coldea_fphy.2020}. Generally, $v_{\rm F}^{(xy)}$ in FeSe is smaller than $v_{\rm F}^{(xy)}$ in \BFCA for instance \cite{YiM:2015} in agreement with  theoretical predictions \cite{Yin:2011,Stadler:2015,Skornyakov:2017}. Thus FeSe is close to the localization limit, and the two-magnon-like response may result from the rather slow carriers on the $d_{xy}$ band. In contrast, the more itinerant carriers in the pnictides condense into a stripe-like spin density wave (SDW) which becomes manifest in a gap and a coherence peak \cite{Kretzschmar:2016,Baum:2019}.

In the energy region $\Omega<200$\,\wn extra intensity is observed for low temperatures. In FeSe it becomes clearly visible below 200\,K and fills the spectral gap below the magnon at 500\,\wn. Below approximately 100\,K an isolated peak may be observed for $x=0$ which continuously softens above the structural transformation at $\Ts=90$\,K, fades away below \Ts, and almost vanishes at 21\,K \cite{Baum:2019,chibani2021npj}. The line shape and the temperature dependence above \Ts can be described quantitatively in terms of critical fluctuations in a similar fashion as in \BFCA \cite{Kretzschmar:2016, Karahasanovic2015_PRB92_075134,Baum:2019}. For increasing doping, this extra intensity starts to develop at lower temperatures. However, remarkably enough the temperature where this peak's intensity is maximal, \Tfm, always coincides with the respective transition temperature \Ts$(x)$. At $x=0.23$ the fluctuation response gains intensity down to the lowest accessible temperatures as presented in the inset of Fig.~\ref{fig:fluct}. Thus, the phase transition line of the nematic phase can also be tracked by the Raman response. For $x=0.33$ (the next available doping level) the fluctuation response cannot be observed any further. Concomitantly, the two-magnon excitation at 500\,\wn becomes unobservable. The most likely explanation of this coincidence is that the two phenomena have the same origin and result from spin excitations. However, there is no consensus on that view in the literature, and Zhang \etal \cite{ZhangWL:2021} and Chibani \etal \cite{Chibani2021} interpret the same experimental observation in terms of quadrupolar charge fluctuations. Yet, one certainly has to answer the question as to why the fluctuations are not found in the simulations \cite{Ruiz:2019}.

Most probably, the length scale the simulations can deal with limits the applicability of the Exact Diagonalization method. Since it was intended to study the temperature dependence the cluster had to be sufficiently small ($4\times4$) to keep the time for the simulations finite \cite{Ruiz:2019}. For the two-magnon excitations, the $4\times4$ cluster is sufficient because only nearest neighbor spins are important. However, close to the transition the correlation length of fluctuations diverges making them inaccessible for the small clusters tractable numerically. Actually, well above the transition there is a shoulder on the low-energy side of the two-magnon peak which may be associated with the fluctuations but the shoulder is lost close to \Ts. Thus, although there are experimental arguments in favor of spin fluctuations at low energy there is no theoretical support for this conjecture.

The last question  we wish to address concerns the origin of possible local or quasi-local spin order in \FESES for $x<0.3$. It was observed a while ago that the width of the various bands derived from the orbitals close to the Fermi surface varies by approximately a factor of three or more. There are itinerant $yz$ and $xz$ bands crossing the Fermi surface at $E_{F}$  and a weakly dispersing $xy$ band just below $E_{F}$ \cite{YiM:2015} on which the nearly localized spins may reside  \cite{Yin:2011,Skornyakov:2017}. It is an important question to which extent the fluctuations at low energy are related to these spins. As a matter of fact, Ba122 displays itinerant SDW magnetism as manifested by a gap and a coherence peak along with fluctuations \cite{Kretzschmar:2016,Baum:2019} whereas FeSe exhibits signatures of local spins and also fluctuations at low energies as shown here.  
In \FESES both phenomena disappear together above $x>0.23$.

%\begin{equation}
  %\Gamma_\mathrm{L}(T) = \Gamma_{\mathrm{L},0} \left(1 + \frac{2    \lambda_{\rm{ph-ph}}}{\exp\left(\frac{\hbar \omega_0}{2 k_{\rm B} T}\right)-1}\right).
%\label{eq:FWHM_fitfunc}
%\end{equation}

%
%\begin{eqnarray}
  %\omega(T) = \omega_0\left[\frac{}{}\right.\!\!1
                             %&-& \gamma  \frac{V(T)-V_0}{V_0} \nonumber \\
                             %&-&\!\left.\left(\!\frac{\Gamma_{\mathrm{L},0}}
                             %{\sqrt{2}\omega_0}\!\right)^{\!2}\!\! \left(1 \!+\!
                             %\frac{4 \lambda_{\rm{ph-ph}}}{\exp\left(\frac{\hbar \omega_0}
                             %{2 k_{\rm B} T}\right) \!-\!1}\right)\right]\!.
%\label{eq:energy_fitfunc}
%\end{eqnarray}

\section{Conclusion}
\label{sec:conclusion}
Raman results covering the entire substitution range $0\le x\le1$ in \FESES were presented. The main goal was the study of the physics around the QCP where the nematic instability approached zero transition temperature in the range $0.16\le x \le 0.23$. We find a striking signature of this transition in both the phonon and the electronic spectra. Whereas the \Blg phonon varies continuously with S substitution the \Alg phonon and all structures in the $xx$ spectra show a discontinuity above $x=0.23$. Similarly, the electronic spectra dominated by spin excitations change abruptly here. Both the two-magnon excitations and the low-energy fluctuations disappear. We argue that they are interrelated. Since we could not observe gap excitation for $x>0$, statements about the evolution of the superconducting pairing are currently not possible. Another issue is the exact position of the quantum critical transition and its sharpness.

\section*{Acknowledgement}
We acknowledge valuable discussions with T. B\"ohm and D. Jost. The authors acknowledge funding provided by the Institute of Physics Belgrade through the grant by the Ministry of Education, Science and Technological Development of the Republic of Serbia and and SASA project No. F-134. The work was supported by the Science Fund of the Republic of Serbia, PROMIS, No. 6062656, StrainedFeSC, and by Research Foundation-Flanders (FWO). Further support came from the German research foundation (DFG) via projects Ha2071/8-1, Ha2071/12-1 and 107745057 – TRR 80 and from the DAAD via the project-related personal exchange program PPP with Serbia grant-no. 57449106. J.B. acknowledges support of a postdoctoral fellowship of the FWO, and of the Erasmus+ program for staff mobility and training (KA107, 2018) for a research stay at the Institute of Physics Belgrade, during which part of the work was carried out. The computational resources and services used for the first-principles calculations in this work were provided by the VSC (Flemish Supercomputer Center), funded by the FWO and the Flemish Government -- department EWI. Work at Brookhaven is supported by the U.S. DOE under Contract No. DESC0012704. 

%\section*{Author contributions}

%A.B.,  conceived exp, performed exp., analyzed and discussed data, wrote paper \\
%A.M. conceived exp, performed exp., analyzed and discussed data, wrote paper \\
%N.L. conceived exp, performed exp., analyzed and discussed data, wrote paper \\
%M.M.R. calculated phonon dispersion and PDOS \\
%B.N. calculated MGPT \\
%M.M. performed exp., analyzed and discussed data \\
%Z.I.M. performed exp. \\
%M.S. performed exp. \\
%M.G.B. performed ellipsometric measurements\\
%N.S. performed exp. \\
%M.O. performed SQUID measurements \\
%A.W. synthesized and characterized the samples \\
%C.P. synthesized and characterized the samples \\
%Z.V.P. analyzed and discussed data, wrote paper \\
%R.H. conceived exp, analyzed and discussed data, wrote paper \\
%All authors commented on the manuscript.

\bibliography{literatur,literatureR3,References}

\begin{thebibliography}{39}%
\makeatletter
\providecommand \@ifxundefined [1]{%
 \@ifx{#1\undefined}
}%
\providecommand \@ifnum [1]{%
 \ifnum #1\expandafter \@firstoftwo
 \else \expandafter \@secondoftwo
 \fi
}%
\providecommand \@ifx [1]{%
 \ifx #1\expandafter \@firstoftwo
 \else \expandafter \@secondoftwo
 \fi
}%
\providecommand \natexlab [1]{#1}%
\providecommand \enquote  [1]{``#1''}%
\providecommand \bibnamefont  [1]{#1}%
\providecommand \bibfnamefont [1]{#1}%
\providecommand \citenamefont [1]{#1}%
\providecommand \href@noop [0]{\@secondoftwo}%
\providecommand \href [0]{\begingroup \@sanitize@url \@href}%
\providecommand \@href[1]{\@@startlink{#1}\@@href}%
\providecommand \@@href[1]{\endgroup#1\@@endlink}%
\providecommand \@sanitize@url [0]{\catcode `\\12\catcode `\$12\catcode
  `\&12\catcode `\#12\catcode `\^12\catcode `\_12\catcode `\%12\relax}%
\providecommand \@@startlink[1]{}%
\providecommand \@@endlink[0]{}%
\providecommand \url  [0]{\begingroup\@sanitize@url \@url }%
\providecommand \@url [1]{\endgroup\@href {#1}{\urlprefix }}%
\providecommand \urlprefix  [0]{URL }%
\providecommand \Eprint [0]{\href }%
\providecommand \doibase [0]{https://doi.org/}%
\providecommand \selectlanguage [0]{\@gobble}%
\providecommand \bibinfo  [0]{\@secondoftwo}%
\providecommand \bibfield  [0]{\@secondoftwo}%
\providecommand \translation [1]{[#1]}%
\providecommand \BibitemOpen [0]{}%
\providecommand \bibitemStop [0]{}%
\providecommand \bibitemNoStop [0]{.\EOS\space}%
\providecommand \EOS [0]{\spacefactor3000\relax}%
\providecommand \BibitemShut  [1]{\csname bibitem#1\endcsname}%
\let\auto@bib@innerbib\@empty
%</preamble>
\bibitem [{\citenamefont {Scalapino}(2012)}]{Scalapino:2012}%
  \BibitemOpen
  \bibfield  {author} {\bibinfo {author} {\bibfnamefont {D.~J.}\ \bibnamefont
  {Scalapino}},\ }\bibfield  {title} {\bibinfo {title} {A common thread: The
  pairing interaction for unconventional superconductors},\ }\href
  {https://doi.org/10.1103/RevModPhys.84.1383} {\bibfield  {journal} {\bibinfo
  {journal} {Rev. Mod. Phys.}\ }\textbf {\bibinfo {volume} {84}},\ \bibinfo
  {pages} {1383} (\bibinfo {year} {2012})}\BibitemShut {NoStop}%
\bibitem [{\citenamefont {Fradkin}\ \emph {et~al.}(2015)\citenamefont
  {Fradkin}, \citenamefont {Kivelson},\ and\ \citenamefont
  {Tranquada}}]{Fradkin:2015}%
  \BibitemOpen
  \bibfield  {author} {\bibinfo {author} {\bibfnamefont {E.}~\bibnamefont
  {Fradkin}}, \bibinfo {author} {\bibfnamefont {S.~A.}\ \bibnamefont
  {Kivelson}},\ and\ \bibinfo {author} {\bibfnamefont {J.~M.}\ \bibnamefont
  {Tranquada}},\ }\bibfield  {title} {\bibinfo {title} {\textit{Colloquium}:
  Theory of intertwined orders in high temperature superconductors},\ }\href
  {https://doi.org/10.1103/RevModPhys.87.457} {\bibfield  {journal} {\bibinfo
  {journal} {Rev. Mod. Phys.}\ }\textbf {\bibinfo {volume} {87}},\ \bibinfo
  {pages} {457} (\bibinfo {year} {2015})}\BibitemShut {NoStop}%
\bibitem [{\citenamefont {Lederer}\ \emph {et~al.}(2015)\citenamefont
  {Lederer}, \citenamefont {Schattner}, \citenamefont {Berg},\ and\
  \citenamefont {Kivelson}}]{Lederer:2015}%
  \BibitemOpen
  \bibfield  {author} {\bibinfo {author} {\bibfnamefont {S.}~\bibnamefont
  {Lederer}}, \bibinfo {author} {\bibfnamefont {Y.}~\bibnamefont {Schattner}},
  \bibinfo {author} {\bibfnamefont {E.}~\bibnamefont {Berg}},\ and\ \bibinfo
  {author} {\bibfnamefont {S.~A.}\ \bibnamefont {Kivelson}},\ }\bibfield
  {title} {\bibinfo {title} {Enhancement of superconductivity near a nematic
  quantum critical point},\ }\href
  {https://doi.org/10.1103/PhysRevLett.114.097001} {\bibfield  {journal}
  {\bibinfo  {journal} {Phys. Rev. Lett.}\ }\textbf {\bibinfo {volume} {114}},\
  \bibinfo {pages} {097001} (\bibinfo {year} {2015})}\BibitemShut {NoStop}%
\bibitem [{\citenamefont {B\"ohmer}\ \emph {et~al.}(2015)\citenamefont
  {B\"ohmer}, \citenamefont {Arai}, \citenamefont {Hardy}, \citenamefont
  {Hattori}, \citenamefont {Iye}, \citenamefont {Wolf}, \citenamefont
  {L\"ohneysen}, \citenamefont {Ishida},\ and\ \citenamefont
  {Meingast}}]{Boehmer2015_PRL114_027001}%
  \BibitemOpen
  \bibfield  {author} {\bibinfo {author} {\bibfnamefont {A.~E.}\ \bibnamefont
  {B\"ohmer}}, \bibinfo {author} {\bibfnamefont {T.}~\bibnamefont {Arai}},
  \bibinfo {author} {\bibfnamefont {F.}~\bibnamefont {Hardy}}, \bibinfo
  {author} {\bibfnamefont {T.}~\bibnamefont {Hattori}}, \bibinfo {author}
  {\bibfnamefont {T.}~\bibnamefont {Iye}}, \bibinfo {author} {\bibfnamefont
  {T.}~\bibnamefont {Wolf}}, \bibinfo {author} {\bibfnamefont {H.~v.}\
  \bibnamefont {L\"ohneysen}}, \bibinfo {author} {\bibfnamefont
  {K.}~\bibnamefont {Ishida}},\ and\ \bibinfo {author} {\bibfnamefont
  {C.}~\bibnamefont {Meingast}},\ }\bibfield  {title} {\bibinfo {title}
  {{Origin of the Tetragonal-to-Orthorhombic Phase Transition in FeSe: A
  Combined Thermodynamic and NMR Study of Nematicity}},\ }\href
  {https://doi.org/10.1103/PhysRevLett.114.027001} {\bibfield  {journal}
  {\bibinfo  {journal} {Phys. Rev. Lett.}\ }\textbf {\bibinfo {volume} {114}},\
  \bibinfo {pages} {027001} (\bibinfo {year} {2015})}\BibitemShut {NoStop}%
\bibitem [{\citenamefont {McQueen}\ \emph {et~al.}(2009)\citenamefont
  {McQueen}, \citenamefont {Williams}, \citenamefont {Stephens}, \citenamefont
  {Tao}, \citenamefont {Zhu}, \citenamefont {Ksenofontov}, \citenamefont
  {Casper}, \citenamefont {Felser},\ and\ \citenamefont
  {Cava}}]{McQueen2009_PRL103_057002}%
  \BibitemOpen
  \bibfield  {author} {\bibinfo {author} {\bibfnamefont {T.~M.}\ \bibnamefont
  {McQueen}}, \bibinfo {author} {\bibfnamefont {A.~J.}\ \bibnamefont
  {Williams}}, \bibinfo {author} {\bibfnamefont {P.~W.}\ \bibnamefont
  {Stephens}}, \bibinfo {author} {\bibfnamefont {J.}~\bibnamefont {Tao}},
  \bibinfo {author} {\bibfnamefont {Y.}~\bibnamefont {Zhu}}, \bibinfo {author}
  {\bibfnamefont {V.}~\bibnamefont {Ksenofontov}}, \bibinfo {author}
  {\bibfnamefont {F.}~\bibnamefont {Casper}}, \bibinfo {author} {\bibfnamefont
  {C.}~\bibnamefont {Felser}},\ and\ \bibinfo {author} {\bibfnamefont {R.~J.}\
  \bibnamefont {Cava}},\ }\bibfield  {title} {\bibinfo {title}
  {{Tetragonal-to-Orthorhombic Structural Phase Transition at 90 K in the
  Superconductor Fe$_{1.01}$Se}},\ }\href
  {https://doi.org/10.1103/PhysRevLett.103.057002} {\bibfield  {journal}
  {\bibinfo  {journal} {Phys. Rev. Lett.}\ }\textbf {\bibinfo {volume} {103}},\
  \bibinfo {pages} {057002} (\bibinfo {year} {2009})}\BibitemShut {NoStop}%
\bibitem [{\citenamefont {Watson}\ \emph {et~al.}(2015)\citenamefont {Watson},
  \citenamefont {Kim}, \citenamefont {Haghighirad}, \citenamefont {Davies},
  \citenamefont {McCollam}, \citenamefont {Narayanan}, \citenamefont {Blake},
  \citenamefont {Chen}, \citenamefont {Ghannadzadeh}, \citenamefont
  {Schofield}, \citenamefont {Hoesch}, \citenamefont {Meingast}, \citenamefont
  {Wolf},\ and\ \citenamefont {Coldea}}]{Watson2015_PRB91_155106}%
  \BibitemOpen
  \bibfield  {author} {\bibinfo {author} {\bibfnamefont {M.~D.}\ \bibnamefont
  {Watson}}, \bibinfo {author} {\bibfnamefont {T.~K.}\ \bibnamefont {Kim}},
  \bibinfo {author} {\bibfnamefont {A.~A.}\ \bibnamefont {Haghighirad}},
  \bibinfo {author} {\bibfnamefont {N.~R.}\ \bibnamefont {Davies}}, \bibinfo
  {author} {\bibfnamefont {A.}~\bibnamefont {McCollam}}, \bibinfo {author}
  {\bibfnamefont {A.}~\bibnamefont {Narayanan}}, \bibinfo {author}
  {\bibfnamefont {S.~F.}\ \bibnamefont {Blake}}, \bibinfo {author}
  {\bibfnamefont {Y.~L.}\ \bibnamefont {Chen}}, \bibinfo {author}
  {\bibfnamefont {S.}~\bibnamefont {Ghannadzadeh}}, \bibinfo {author}
  {\bibfnamefont {A.~J.}\ \bibnamefont {Schofield}}, \bibinfo {author}
  {\bibfnamefont {M.}~\bibnamefont {Hoesch}}, \bibinfo {author} {\bibfnamefont
  {C.}~\bibnamefont {Meingast}}, \bibinfo {author} {\bibfnamefont
  {T.}~\bibnamefont {Wolf}},\ and\ \bibinfo {author} {\bibfnamefont {A.~I.}\
  \bibnamefont {Coldea}},\ }\bibfield  {title} {\bibinfo {title} {{Emergence of
  the nematic electronic state in FeSe}},\ }\href
  {https://doi.org/10.1103/PhysRevB.91.155106} {\bibfield  {journal} {\bibinfo
  {journal} {Phys. Rev. B}\ }\textbf {\bibinfo {volume} {91}},\ \bibinfo
  {pages} {155106} (\bibinfo {year} {2015})}\BibitemShut {NoStop}%
\bibitem [{\citenamefont {Hsu}\ \emph {et~al.}(2008)\citenamefont {Hsu},
  \citenamefont {Luo}, \citenamefont {Yeh}, \citenamefont {Chen}, \citenamefont
  {Huang}, \citenamefont {Wu}, \citenamefont {Lee}, \citenamefont {Huang},
  \citenamefont {Chu}, \citenamefont {Yan},\ and\ \citenamefont
  {Wu}}]{Hsu2008_PNAS105_14262}%
  \BibitemOpen
  \bibfield  {author} {\bibinfo {author} {\bibfnamefont {F.-C.}\ \bibnamefont
  {Hsu}}, \bibinfo {author} {\bibfnamefont {J.-Y.}\ \bibnamefont {Luo}},
  \bibinfo {author} {\bibfnamefont {K.-W.}\ \bibnamefont {Yeh}}, \bibinfo
  {author} {\bibfnamefont {T.-K.}\ \bibnamefont {Chen}}, \bibinfo {author}
  {\bibfnamefont {T.-W.}\ \bibnamefont {Huang}}, \bibinfo {author}
  {\bibfnamefont {P.~M.}\ \bibnamefont {Wu}}, \bibinfo {author} {\bibfnamefont
  {Y.-C.}\ \bibnamefont {Lee}}, \bibinfo {author} {\bibfnamefont {Y.-L.}\
  \bibnamefont {Huang}}, \bibinfo {author} {\bibfnamefont {Y.-Y.}\ \bibnamefont
  {Chu}}, \bibinfo {author} {\bibfnamefont {D.-C.}\ \bibnamefont {Yan}},\ and\
  \bibinfo {author} {\bibfnamefont {M.-K.}\ \bibnamefont {Wu}},\ }\bibfield
  {title} {\bibinfo {title} {{Superconductivity in the PbO-type structure
  $\alpha$-FeSe}},\ }\href {https://doi.org/10.1073/pnas.0807325105} {\bibfield
   {journal} {\bibinfo  {journal} {Proc. Natl. Acad. Sci.}\ }\textbf {\bibinfo
  {volume} {105}},\ \bibinfo {pages} {14262} (\bibinfo {year}
  {2008})}\BibitemShut {NoStop}%
\bibitem [{\citenamefont {Medvedev}\ \emph {et~al.}(2009)\citenamefont
  {Medvedev}, \citenamefont {McQueen}, \citenamefont {Troyan}, \citenamefont
  {Palasyuk}, \citenamefont {Eremets}, \citenamefont {Cava}, \citenamefont
  {Naghavi}, \citenamefont {Casper}, \citenamefont {Ksenofontov}, \citenamefont
  {Wortmann},\ and\ \citenamefont {Felser}}]{Medvedev2009_NM8_630}%
  \BibitemOpen
  \bibfield  {author} {\bibinfo {author} {\bibfnamefont {S.}~\bibnamefont
  {Medvedev}}, \bibinfo {author} {\bibfnamefont {T.~M.}\ \bibnamefont
  {McQueen}}, \bibinfo {author} {\bibfnamefont {I.~A.}\ \bibnamefont {Troyan}},
  \bibinfo {author} {\bibfnamefont {T.}~\bibnamefont {Palasyuk}}, \bibinfo
  {author} {\bibfnamefont {M.~I.}\ \bibnamefont {Eremets}}, \bibinfo {author}
  {\bibfnamefont {R.~J.}\ \bibnamefont {Cava}}, \bibinfo {author}
  {\bibfnamefont {S.}~\bibnamefont {Naghavi}}, \bibinfo {author} {\bibfnamefont
  {F.}~\bibnamefont {Casper}}, \bibinfo {author} {\bibfnamefont
  {V.}~\bibnamefont {Ksenofontov}}, \bibinfo {author} {\bibfnamefont
  {G.}~\bibnamefont {Wortmann}},\ and\ \bibinfo {author} {\bibfnamefont
  {C.}~\bibnamefont {Felser}},\ }\bibfield  {title} {\bibinfo {title}
  {{Electronic and magnetic phase diagram of $\beta$-Fe$_{1.01}$Se with
  superconductivity at 36.7K under pressure}},\ }\href
  {https://doi.org/10.1038/NMAT2491} {\bibfield  {journal} {\bibinfo  {journal}
  {Nat. Mater.}\ }\textbf {\bibinfo {volume} {8}},\ \bibinfo {pages} {630}
  (\bibinfo {year} {2009})}\BibitemShut {NoStop}%
\bibitem [{\citenamefont {Sato}\ \emph {et~al.}(2018)\citenamefont {Sato},
  \citenamefont {Kasahara}, \citenamefont {Taniguchi}, \citenamefont {Xing},
  \citenamefont {Kasahara}, \citenamefont {Tokiwa}, \citenamefont {Yamakawa},
  \citenamefont {Kontani}, \citenamefont {Shibauchi},\ and\ \citenamefont
  {Matsuda}}]{Sato1227}%
  \BibitemOpen
  \bibfield  {author} {\bibinfo {author} {\bibfnamefont {Y.}~\bibnamefont
  {Sato}}, \bibinfo {author} {\bibfnamefont {S.}~\bibnamefont {Kasahara}},
  \bibinfo {author} {\bibfnamefont {T.}~\bibnamefont {Taniguchi}}, \bibinfo
  {author} {\bibfnamefont {X.}~\bibnamefont {Xing}}, \bibinfo {author}
  {\bibfnamefont {Y.}~\bibnamefont {Kasahara}}, \bibinfo {author}
  {\bibfnamefont {Y.}~\bibnamefont {Tokiwa}}, \bibinfo {author} {\bibfnamefont
  {Y.}~\bibnamefont {Yamakawa}}, \bibinfo {author} {\bibfnamefont
  {H.}~\bibnamefont {Kontani}}, \bibinfo {author} {\bibfnamefont
  {T.}~\bibnamefont {Shibauchi}},\ and\ \bibinfo {author} {\bibfnamefont
  {Y.}~\bibnamefont {Matsuda}},\ }\bibfield  {title} {\bibinfo {title} {Abrupt
  change of the superconducting gap structure at the nematic critical point in
  {FeSe$_{1-x}$S$_x$}},\ }\href {https://doi.org/10.1073/pnas.1717331115}
  {\bibfield  {journal} {\bibinfo  {journal} {Proceedings of the National
  Academy of Sciences}\ }\textbf {\bibinfo {volume} {115}},\ \bibinfo {pages}
  {1227} (\bibinfo {year} {2018})}\BibitemShut {NoStop}%
\bibitem [{\citenamefont {Lai}\ \emph {et~al.}(2015)\citenamefont {Lai},
  \citenamefont {Zhang}, \citenamefont {Wang}, \citenamefont {Wang},
  \citenamefont {Zhang}, \citenamefont {Lin},\ and\ \citenamefont
  {Huang}}]{Lai2015_JACS137_10148}%
  \BibitemOpen
  \bibfield  {author} {\bibinfo {author} {\bibfnamefont {X.}~\bibnamefont
  {Lai}}, \bibinfo {author} {\bibfnamefont {H.}~\bibnamefont {Zhang}}, \bibinfo
  {author} {\bibfnamefont {Y.}~\bibnamefont {Wang}}, \bibinfo {author}
  {\bibfnamefont {X.}~\bibnamefont {Wang}}, \bibinfo {author} {\bibfnamefont
  {X.}~\bibnamefont {Zhang}}, \bibinfo {author} {\bibfnamefont
  {J.}~\bibnamefont {Lin}},\ and\ \bibinfo {author} {\bibfnamefont
  {F.}~\bibnamefont {Huang}},\ }\bibfield  {title} {\bibinfo {title}
  {{Observation of Superconductivity in Tetragonal FeS}},\ }\href
  {https://doi.org/10.1021/jacs.5b06687} {\bibfield  {journal} {\bibinfo
  {journal} {J. Am. Chem. Soc.}\ }\textbf {\bibinfo {volume} {137}},\ \bibinfo
  {pages} {10148} (\bibinfo {year} {2015})}\BibitemShut {NoStop}%
\bibitem [{\citenamefont {Pachmayr}\ \emph {et~al.}(2016)\citenamefont
  {Pachmayr}, \citenamefont {Fehn},\ and\ \citenamefont
  {Johrendt}}]{Pachmayr2016_CC52_194}%
  \BibitemOpen
  \bibfield  {author} {\bibinfo {author} {\bibfnamefont {U.}~\bibnamefont
  {Pachmayr}}, \bibinfo {author} {\bibfnamefont {N.}~\bibnamefont {Fehn}},\
  and\ \bibinfo {author} {\bibfnamefont {D.}~\bibnamefont {Johrendt}},\
  }\bibfield  {title} {\bibinfo {title} {{Structural transition and
  superconductivity in hydrothermally synthesized Fe$X$ ($X$ = S, Se)}},\
  }\href {https://doi.org/10.1039/C5CC07739G} {\bibfield  {journal} {\bibinfo
  {journal} {Chem. Commun.}\ }\textbf {\bibinfo {volume} {52}},\ \bibinfo
  {pages} {194} (\bibinfo {year} {2016})}\BibitemShut {NoStop}%
\bibitem [{\citenamefont {Lazarevi{\'{c}}}\ and\ \citenamefont
  {Hackl}(2020)}]{Lazarevic_2020}%
  \BibitemOpen
  \bibfield  {author} {\bibinfo {author} {\bibfnamefont {N.}~\bibnamefont
  {Lazarevi{\'{c}}}}\ and\ \bibinfo {author} {\bibfnamefont {R.}~\bibnamefont
  {Hackl}},\ }\bibfield  {title} {\bibinfo {title} {{Fluctuations and pairing
  in Fe-based superconductors: light scattering experiments}},\ }\href
  {https://doi.org/10.1088/1361-648x/ab8849} {\bibfield  {journal} {\bibinfo
  {journal} {Journal of Physics: Condensed Matter}\ }\textbf {\bibinfo {volume}
  {32}},\ \bibinfo {pages} {413001} (\bibinfo {year} {2020})}\BibitemShut
  {NoStop}%
\bibitem [{\citenamefont {Holenstein}\ \emph {et~al.}(2016)\citenamefont
  {Holenstein}, \citenamefont {Pachmayr}, \citenamefont {Guguchia},
  \citenamefont {Kamusella}, \citenamefont {Khasanov}, \citenamefont {Amato},
  \citenamefont {Baines}, \citenamefont {Klauss}, \citenamefont {Morenzoni},
  \citenamefont {Johrendt},\ and\ \citenamefont
  {Luetkens}}]{Holenstein2016_PRB93_140506}%
  \BibitemOpen
  \bibfield  {author} {\bibinfo {author} {\bibfnamefont {S.}~\bibnamefont
  {Holenstein}}, \bibinfo {author} {\bibfnamefont {U.}~\bibnamefont
  {Pachmayr}}, \bibinfo {author} {\bibfnamefont {Z.}~\bibnamefont {Guguchia}},
  \bibinfo {author} {\bibfnamefont {S.}~\bibnamefont {Kamusella}}, \bibinfo
  {author} {\bibfnamefont {R.}~\bibnamefont {Khasanov}}, \bibinfo {author}
  {\bibfnamefont {A.}~\bibnamefont {Amato}}, \bibinfo {author} {\bibfnamefont
  {C.}~\bibnamefont {Baines}}, \bibinfo {author} {\bibfnamefont {H.-H.}\
  \bibnamefont {Klauss}}, \bibinfo {author} {\bibfnamefont {E.}~\bibnamefont
  {Morenzoni}}, \bibinfo {author} {\bibfnamefont {D.}~\bibnamefont
  {Johrendt}},\ and\ \bibinfo {author} {\bibfnamefont {H.}~\bibnamefont
  {Luetkens}},\ }\bibfield  {title} {\bibinfo {title} {{Coexistence of
  low-moment magnetism and superconductivity in tetragonal FeS and suppression
  of T$_\mathrm{c}$ under pressure}},\ }\href
  {https://doi.org/10.1103/PhysRevB.93.140506} {\bibfield  {journal} {\bibinfo
  {journal} {Phys. Rev. B}\ }\textbf {\bibinfo {volume} {93}},\ \bibinfo
  {pages} {140506} (\bibinfo {year} {2016})}\BibitemShut {NoStop}%
\bibitem [{\citenamefont {Kirschner}\ \emph {et~al.}(2016)\citenamefont
  {Kirschner}, \citenamefont {Lang}, \citenamefont {Topping}, \citenamefont
  {Baker}, \citenamefont {Pratt}, \citenamefont {Wright}, \citenamefont
  {Woodruff}, \citenamefont {Clarke},\ and\ \citenamefont
  {Blundell}}]{Kirschner2016_PRB94_134509}%
  \BibitemOpen
  \bibfield  {author} {\bibinfo {author} {\bibfnamefont {F.~K.~K.}\
  \bibnamefont {Kirschner}}, \bibinfo {author} {\bibfnamefont {F.}~\bibnamefont
  {Lang}}, \bibinfo {author} {\bibfnamefont {C.~V.}\ \bibnamefont {Topping}},
  \bibinfo {author} {\bibfnamefont {P.~J.}\ \bibnamefont {Baker}}, \bibinfo
  {author} {\bibfnamefont {F.~L.}\ \bibnamefont {Pratt}}, \bibinfo {author}
  {\bibfnamefont {S.~E.}\ \bibnamefont {Wright}}, \bibinfo {author}
  {\bibfnamefont {D.~N.}\ \bibnamefont {Woodruff}}, \bibinfo {author}
  {\bibfnamefont {S.~J.}\ \bibnamefont {Clarke}},\ and\ \bibinfo {author}
  {\bibfnamefont {S.~J.}\ \bibnamefont {Blundell}},\ }\bibfield  {title}
  {\bibinfo {title} {{Robustness of superconductivity to competing magnetic
  phases in tetragonal FeS}},\ }\href
  {https://doi.org/10.1103/PhysRevB.94.134509} {\bibfield  {journal} {\bibinfo
  {journal} {Phys. Rev. B}\ }\textbf {\bibinfo {volume} {94}},\ \bibinfo
  {pages} {134509} (\bibinfo {year} {2016})}\BibitemShut {NoStop}%
\bibitem [{\citenamefont {Wang}\ \emph {et~al.}(2016)\citenamefont {Wang},
  \citenamefont {Wu}, \citenamefont {Ivanovski}, \citenamefont {Warren},
  \citenamefont {Tian}, \citenamefont {Zhu},\ and\ \citenamefont
  {Petrovic}}]{Wang2016_PRB94_094506}%
  \BibitemOpen
  \bibfield  {author} {\bibinfo {author} {\bibfnamefont {A.}~\bibnamefont
  {Wang}}, \bibinfo {author} {\bibfnamefont {L.}~\bibnamefont {Wu}}, \bibinfo
  {author} {\bibfnamefont {V.~N.}\ \bibnamefont {Ivanovski}}, \bibinfo {author}
  {\bibfnamefont {J.~B.}\ \bibnamefont {Warren}}, \bibinfo {author}
  {\bibfnamefont {J.}~\bibnamefont {Tian}}, \bibinfo {author} {\bibfnamefont
  {Y.}~\bibnamefont {Zhu}},\ and\ \bibinfo {author} {\bibfnamefont
  {C.}~\bibnamefont {Petrovic}},\ }\bibfield  {title} {\bibinfo {title}
  {{Critical current density and vortex pinning in tetragonal
  FeS$_{1-x}$Se$_{x}$ ($x=0,0.06$)}},\ }\href
  {https://doi.org/10.1103/PhysRevB.94.094506} {\bibfield  {journal} {\bibinfo
  {journal} {Phys. Rev. B}\ }\textbf {\bibinfo {volume} {94}},\ \bibinfo
  {pages} {094506} (\bibinfo {year} {2016})}\BibitemShut {NoStop}%
\bibitem [{\citenamefont {Gnezdilov}\ \emph {et~al.}(2013)\citenamefont
  {Gnezdilov}, \citenamefont {Pashkevich}, \citenamefont {Lemmens},
  \citenamefont {Wulferding}, \citenamefont {Shevtsova}, \citenamefont {Gusev},
  \citenamefont {Chareev},\ and\ \citenamefont
  {Vasiliev}}]{Gnezdilov2013_PRB87_144508}%
  \BibitemOpen
  \bibfield  {author} {\bibinfo {author} {\bibfnamefont {V.}~\bibnamefont
  {Gnezdilov}}, \bibinfo {author} {\bibfnamefont {Y.~G.}\ \bibnamefont
  {Pashkevich}}, \bibinfo {author} {\bibfnamefont {P.}~\bibnamefont {Lemmens}},
  \bibinfo {author} {\bibfnamefont {D.}~\bibnamefont {Wulferding}}, \bibinfo
  {author} {\bibfnamefont {T.}~\bibnamefont {Shevtsova}}, \bibinfo {author}
  {\bibfnamefont {A.}~\bibnamefont {Gusev}}, \bibinfo {author} {\bibfnamefont
  {D.}~\bibnamefont {Chareev}},\ and\ \bibinfo {author} {\bibfnamefont
  {A.}~\bibnamefont {Vasiliev}},\ }\bibfield  {title} {\bibinfo {title}
  {{Interplay between lattice and spin states degree of freedom in the FeSe
  superconductor: Dynamic spin state instabilities}},\ }\href
  {https://doi.org/10.1103/PhysRevB.87.144508} {\bibfield  {journal} {\bibinfo
  {journal} {Phys. Rev. B}\ }\textbf {\bibinfo {volume} {87}},\ \bibinfo
  {pages} {144508} (\bibinfo {year} {2013})}\BibitemShut {NoStop}%
\bibitem [{\citenamefont {Baum}\ \emph {et~al.}(2018)\citenamefont {Baum},
  \citenamefont {Milosavljevi\ifmmode~\acute{c}\else \'{c}\fi{}}, \citenamefont
  {Lazarevi\ifmmode~\acute{c}\else \'{c}\fi{}}, \citenamefont
  {Radonji\ifmmode~\acute{c}\else \'{c}\fi{}}, \citenamefont
  {Nikoli\ifmmode~\acute{c}\else \'{c}\fi{}}, \citenamefont {Mitschek},
  \citenamefont {Maranloo}, \citenamefont {\ifmmode \check{S}\else
  \v{S}\fi{}\ifmmode \acute{c}\else \'{c}\fi{}epanovi\ifmmode~\acute{c}\else
  \'{c}\fi{}}, \citenamefont {Gruji\ifmmode \acute{c}\else
  \'{c}\fi{}-Broj\ifmmode~\check{c}\else \v{c}\fi{}in}, \citenamefont
  {Stojilovi\ifmmode~\acute{c}\else \'{c}\fi{}}, \citenamefont {Opel},
  \citenamefont {Wang}, \citenamefont {Petrovic}, \citenamefont
  {Popovi\ifmmode~\acute{c}\else \'{c}\fi{}},\ and\ \citenamefont
  {Hackl}}]{Baum:2018}%
  \BibitemOpen
  \bibfield  {author} {\bibinfo {author} {\bibfnamefont {A.}~\bibnamefont
  {Baum}}, \bibinfo {author} {\bibfnamefont {A.}~\bibnamefont
  {Milosavljevi\ifmmode~\acute{c}\else \'{c}\fi{}}}, \bibinfo {author}
  {\bibfnamefont {N.}~\bibnamefont {Lazarevi\ifmmode~\acute{c}\else
  \'{c}\fi{}}}, \bibinfo {author} {\bibfnamefont {M.~M.}\ \bibnamefont
  {Radonji\ifmmode~\acute{c}\else \'{c}\fi{}}}, \bibinfo {author}
  {\bibfnamefont {B.}~\bibnamefont {Nikoli\ifmmode~\acute{c}\else \'{c}\fi{}}},
  \bibinfo {author} {\bibfnamefont {M.}~\bibnamefont {Mitschek}}, \bibinfo
  {author} {\bibfnamefont {Z.~I.}\ \bibnamefont {Maranloo}}, \bibinfo {author}
  {\bibfnamefont {M.}~\bibnamefont {\ifmmode \check{S}\else \v{S}\fi{}\ifmmode
  \acute{c}\else \'{c}\fi{}epanovi\ifmmode~\acute{c}\else \'{c}\fi{}}},
  \bibinfo {author} {\bibfnamefont {M.}~\bibnamefont {Gruji\ifmmode
  \acute{c}\else \'{c}\fi{}-Broj\ifmmode~\check{c}\else \v{c}\fi{}in}},
  \bibinfo {author} {\bibfnamefont {N.}~\bibnamefont
  {Stojilovi\ifmmode~\acute{c}\else \'{c}\fi{}}}, \bibinfo {author}
  {\bibfnamefont {M.}~\bibnamefont {Opel}}, \bibinfo {author} {\bibfnamefont
  {A.}~\bibnamefont {Wang}}, \bibinfo {author} {\bibfnamefont {C.}~\bibnamefont
  {Petrovic}}, \bibinfo {author} {\bibfnamefont {Z.~V.}\ \bibnamefont
  {Popovi\ifmmode~\acute{c}\else \'{c}\fi{}}},\ and\ \bibinfo {author}
  {\bibfnamefont {R.}~\bibnamefont {Hackl}},\ }\bibfield  {title} {\bibinfo
  {title} {{Phonon anomalies in FeS}},\ }\href
  {https://doi.org/10.1103/PhysRevB.97.054306} {\bibfield  {journal} {\bibinfo
  {journal} {Phys. Rev. B}\ }\textbf {\bibinfo {volume} {97}},\ \bibinfo
  {pages} {054306} (\bibinfo {year} {2018})}\BibitemShut {NoStop}%
\bibitem [{\citenamefont {{Wang}}\ \emph {et~al.}(2020)\citenamefont {{Wang}},
  \citenamefont {{Stavitski}}, \citenamefont {{Naamneh}}, \citenamefont
  {{Ivanovski}}, \citenamefont {{Abeykoon}}, \citenamefont {{Milosavljevic}},
  \citenamefont {{Brito}}, \citenamefont {{Baum}}, \citenamefont {{Jandke}},
  \citenamefont {{Du}}, \citenamefont {{Lazarevic}}, \citenamefont {{Liu}},
  \citenamefont {{Plumb}}, \citenamefont {{Kotliar}}, \citenamefont {{Hackl}},
  \citenamefont {{Popovic}}, \citenamefont {{Radovic}}, \citenamefont
  {{Attenkofer}},\ and\ \citenamefont {{Petrovic}}}]{WangA:2020}%
  \BibitemOpen
  \bibfield  {author} {\bibinfo {author} {\bibfnamefont {A.}~\bibnamefont
  {{Wang}}}, \bibinfo {author} {\bibfnamefont {E.}~\bibnamefont {{Stavitski}}},
  \bibinfo {author} {\bibfnamefont {M.}~\bibnamefont {{Naamneh}}}, \bibinfo
  {author} {\bibfnamefont {V.~N.}\ \bibnamefont {{Ivanovski}}}, \bibinfo
  {author} {\bibfnamefont {M.}~\bibnamefont {{Abeykoon}}}, \bibinfo {author}
  {\bibfnamefont {A.}~\bibnamefont {{Milosavljevic}}}, \bibinfo {author}
  {\bibfnamefont {W.~H.}\ \bibnamefont {{Brito}}}, \bibinfo {author}
  {\bibfnamefont {A.}~\bibnamefont {{Baum}}}, \bibinfo {author} {\bibfnamefont
  {J.}~\bibnamefont {{Jandke}}}, \bibinfo {author} {\bibfnamefont
  {Q.}~\bibnamefont {{Du}}}, \bibinfo {author} {\bibfnamefont {N.}~\bibnamefont
  {{Lazarevic}}}, \bibinfo {author} {\bibfnamefont {Y.}~\bibnamefont {{Liu}}},
  \bibinfo {author} {\bibfnamefont {N.~C.}\ \bibnamefont {{Plumb}}}, \bibinfo
  {author} {\bibfnamefont {G.}~\bibnamefont {{Kotliar}}}, \bibinfo {author}
  {\bibfnamefont {R.}~\bibnamefont {{Hackl}}}, \bibinfo {author} {\bibfnamefont
  {Z.~V.}\ \bibnamefont {{Popovic}}}, \bibinfo {author} {\bibfnamefont
  {M.}~\bibnamefont {{Radovic}}}, \bibinfo {author} {\bibfnamefont
  {K.}~\bibnamefont {{Attenkofer}}},\ and\ \bibinfo {author} {\bibfnamefont
  {C.}~\bibnamefont {{Petrovic}}},\ }\bibfield  {title} {\bibinfo {title}
  {{Superconducting Order from Local Disorder}},\ }\href@noop {} {\bibfield
  {journal} {\bibinfo  {journal} {arXiv e-prints}\ ,\ \bibinfo {eid}
  {arXiv:2009.06623}} (\bibinfo {year} {2020})},\ \Eprint
  {https://arxiv.org/abs/2009.06623} {arXiv:2009.06623 [cond-mat.supr-con]}
  \BibitemShut {NoStop}%
\bibitem [{\citenamefont {Cardona}\ and\ \citenamefont
  {Thewalt}(2005)}]{Cardona:2005}%
  \BibitemOpen
  \bibfield  {author} {\bibinfo {author} {\bibfnamefont {M.}~\bibnamefont
  {Cardona}}\ and\ \bibinfo {author} {\bibfnamefont {M.~L.~W.}\ \bibnamefont
  {Thewalt}},\ }\bibfield  {title} {\bibinfo {title} {Isotope effects on the
  optical spectra of semiconductors},\ }\href
  {https://doi.org/10.1103/RevModPhys.77.1173} {\bibfield  {journal} {\bibinfo
  {journal} {Rev. Mod. Phys.}\ }\textbf {\bibinfo {volume} {77}},\ \bibinfo
  {pages} {1173} (\bibinfo {year} {2005})}\BibitemShut {NoStop}%
\bibitem [{\citenamefont {Turrell}(1972)}]{Turrell:1972}%
  \BibitemOpen
  \bibfield  {author} {\bibinfo {author} {\bibfnamefont {G.}~\bibnamefont
  {Turrell}},\ }\href@noop {} {\emph {\bibinfo {title} {{Infrared and Raman
  spectra of Crystals}}}}\ (\bibinfo  {publisher} {Academic Press inc.},\
  \bibinfo {address} {London and New York},\ \bibinfo {year}
  {1972})\BibitemShut {NoStop}%
\bibitem [{\citenamefont {Mizukami}\ \emph {et~al.}(2021)\citenamefont
  {Mizukami}, \citenamefont {Haze}, \citenamefont {Tanaka}, \citenamefont
  {Matsuura}, \citenamefont {Sano}, \citenamefont {B\"oker}, \citenamefont
  {Eremin}, \citenamefont {Kasahara}, \citenamefont {Matsuda},\ and\
  \citenamefont {Shibauchi}}]{Mitsukami:2021}%
  \BibitemOpen
  \bibfield  {author} {\bibinfo {author} {\bibfnamefont {Y.}~\bibnamefont
  {Mizukami}}, \bibinfo {author} {\bibfnamefont {M.}~\bibnamefont {Haze}},
  \bibinfo {author} {\bibfnamefont {O.}~\bibnamefont {Tanaka}}, \bibinfo
  {author} {\bibfnamefont {K.}~\bibnamefont {Matsuura}}, \bibinfo {author}
  {\bibfnamefont {D.}~\bibnamefont {Sano}}, \bibinfo {author} {\bibfnamefont
  {J.}~\bibnamefont {B\"oker}}, \bibinfo {author} {\bibfnamefont
  {I.}~\bibnamefont {Eremin}}, \bibinfo {author} {\bibfnamefont
  {S.}~\bibnamefont {Kasahara}}, \bibinfo {author} {\bibfnamefont
  {Y.}~\bibnamefont {Matsuda}},\ and\ \bibinfo {author} {\bibfnamefont
  {T.}~\bibnamefont {Shibauchi}},\ }\bibfield  {title} {\bibinfo {title}
  {{Thermodynamics of transition to BCS-BEC crossover superconductivity in
  FeSe$_{1-x}$S$_x$}},\ }\href@noop {} {\bibfield  {journal} {\bibinfo
  {journal} {arXiv e-prints}\ ,\ \bibinfo {eid} {arXiv:2009.06623}} (\bibinfo
  {year} {2021})},\ \Eprint {https://arxiv.org/abs/2105.00739}
  {arXiv:2105.00739 [cond-mat.supr-con]} \BibitemShut {NoStop}%
\bibitem [{\citenamefont {Baum}\ \emph {et~al.}(2019)\citenamefont {Baum},
  \citenamefont {Ruiz}, \citenamefont {Lazarevi\'c}, \citenamefont {Wang},
  \citenamefont {B\"ohm}, \citenamefont {Hosseinian~Ahangharnejhad},
  \citenamefont {Adelmann}, \citenamefont {Wolf}, \citenamefont {Popovi\'c},
  \citenamefont {Moritz}, \citenamefont {Devereaux},\ and\ \citenamefont
  {Hackl}}]{Baum:2019}%
  \BibitemOpen
  \bibfield  {author} {\bibinfo {author} {\bibfnamefont {A.}~\bibnamefont
  {Baum}}, \bibinfo {author} {\bibfnamefont {H.~N.}\ \bibnamefont {Ruiz}},
  \bibinfo {author} {\bibfnamefont {N.}~\bibnamefont {Lazarevi\'c}}, \bibinfo
  {author} {\bibfnamefont {Y.}~\bibnamefont {Wang}}, \bibinfo {author}
  {\bibfnamefont {T.}~\bibnamefont {B\"ohm}}, \bibinfo {author} {\bibfnamefont
  {R.}~\bibnamefont {Hosseinian~Ahangharnejhad}}, \bibinfo {author}
  {\bibfnamefont {P.}~\bibnamefont {Adelmann}}, \bibinfo {author}
  {\bibfnamefont {T.}~\bibnamefont {Wolf}}, \bibinfo {author} {\bibfnamefont
  {Z.~V.}\ \bibnamefont {Popovi\'c}}, \bibinfo {author} {\bibfnamefont
  {B.}~\bibnamefont {Moritz}}, \bibinfo {author} {\bibfnamefont {T.~P.}\
  \bibnamefont {Devereaux}},\ and\ \bibinfo {author} {\bibfnamefont
  {R.}~\bibnamefont {Hackl}},\ }\bibfield  {title} {\bibinfo {title}
  {{Frustrated spin order and stripe fluctuations in FeSe}},\ }\href
  {https://doi.org/10.1038/s42005-019-0107-y} {\bibfield  {journal} {\bibinfo
  {journal} {Commun. Phys.}\ }\textbf {\bibinfo {volume} {2}},\ \bibinfo
  {pages} {14} (\bibinfo {year} {2019})}\BibitemShut {NoStop}%
\bibitem [{\citenamefont {Ruiz}\ \emph {et~al.}(2019)\citenamefont {Ruiz},
  \citenamefont {Wang}, \citenamefont {Moritz}, \citenamefont {Baum},
  \citenamefont {Hackl},\ and\ \citenamefont {Devereaux}}]{Ruiz:2019}%
  \BibitemOpen
  \bibfield  {author} {\bibinfo {author} {\bibfnamefont {H.}~\bibnamefont
  {Ruiz}}, \bibinfo {author} {\bibfnamefont {Y.}~\bibnamefont {Wang}}, \bibinfo
  {author} {\bibfnamefont {B.}~\bibnamefont {Moritz}}, \bibinfo {author}
  {\bibfnamefont {A.}~\bibnamefont {Baum}}, \bibinfo {author} {\bibfnamefont
  {R.}~\bibnamefont {Hackl}},\ and\ \bibinfo {author} {\bibfnamefont {T.~P.}\
  \bibnamefont {Devereaux}},\ }\bibfield  {title} {\bibinfo {title}
  {{Frustrated magnetism from local moments in FeSe}},\ }\href
  {https://doi.org/10.1103/PhysRevB.99.125130} {\bibfield  {journal} {\bibinfo
  {journal} {Phys. Rev. B}\ }\textbf {\bibinfo {volume} {99}},\ \bibinfo
  {pages} {125130} (\bibinfo {year} {2019})}\BibitemShut {NoStop}%
\bibitem [{\citenamefont {Glasbrenner}\ \emph {et~al.}(2015)\citenamefont
  {Glasbrenner}, \citenamefont {Mazin}, \citenamefont {Jeschke}, \citenamefont
  {Hirschfeld}, \citenamefont {Fernandes},\ and\ \citenamefont
  {Valent\'i}}]{Glasbrenner:2015}%
  \BibitemOpen
  \bibfield  {author} {\bibinfo {author} {\bibfnamefont {J.~K.}\ \bibnamefont
  {Glasbrenner}}, \bibinfo {author} {\bibfnamefont {I.~I.}\ \bibnamefont
  {Mazin}}, \bibinfo {author} {\bibfnamefont {H.~O.}\ \bibnamefont {Jeschke}},
  \bibinfo {author} {\bibfnamefont {P.~J.}\ \bibnamefont {Hirschfeld}},
  \bibinfo {author} {\bibfnamefont {R.~M.}\ \bibnamefont {Fernandes}},\ and\
  \bibinfo {author} {\bibfnamefont {R.}~\bibnamefont {Valent\'i}},\ }\bibfield
  {title} {\bibinfo {title} {Effect of magnetic frustration on nematicity and
  superconductivity in iron chalcogenides},\ }\href
  {https://doi.org/10.1038/nphys3434} {\bibfield  {journal} {\bibinfo
  {journal} {Nature Phys.}\ }\textbf {\bibinfo {volume} {11}},\ \bibinfo
  {pages} {953} (\bibinfo {year} {2015})}\BibitemShut {NoStop}%
\bibitem [{\citenamefont {Coldea}(2021)}]{Coldea_fphy.2020}%
  \BibitemOpen
  \bibfield  {author} {\bibinfo {author} {\bibfnamefont {A.~I.}\ \bibnamefont
  {Coldea}},\ }\bibfield  {title} {\bibinfo {title} {{Electronic Nematic States
  Tuned by Isoelectronic Substitution in Bulk FeSe$_{1-x}$S$_x$}},\ }\href
  {https://doi.org/10.3389/fphy.2020.594500} {\bibfield  {journal} {\bibinfo
  {journal} {Frontiers in Physics}\ }\textbf {\bibinfo {volume} {8}},\ \bibinfo
  {pages} {528} (\bibinfo {year} {2021})}\BibitemShut {NoStop}%
\bibitem [{\citenamefont {Yi}\ \emph {et~al.}(2015)\citenamefont {Yi},
  \citenamefont {Liu}, \citenamefont {Zhang}, \citenamefont {Yu}, \citenamefont
  {Zhu}, \citenamefont {Lee}, \citenamefont {Moore}, \citenamefont {Schmitt},
  \citenamefont {Li}, \citenamefont {Riggs}, \citenamefont {Chu}, \citenamefont
  {Lv}, \citenamefont {Hu}, \citenamefont {Hashimoto}, \citenamefont {Mo},
  \citenamefont {Hussain}, \citenamefont {Mao}, \citenamefont {Chu},
  \citenamefont {Fisher}, \citenamefont {Si}, \citenamefont {Shen},\ and\
  \citenamefont {Lu}}]{YiM:2015}%
  \BibitemOpen
  \bibfield  {author} {\bibinfo {author} {\bibfnamefont {M.}~\bibnamefont
  {Yi}}, \bibinfo {author} {\bibfnamefont {Z.-K.}\ \bibnamefont {Liu}},
  \bibinfo {author} {\bibfnamefont {Y.}~\bibnamefont {Zhang}}, \bibinfo
  {author} {\bibfnamefont {R.}~\bibnamefont {Yu}}, \bibinfo {author}
  {\bibfnamefont {J.-X.}\ \bibnamefont {Zhu}}, \bibinfo {author} {\bibfnamefont
  {J.}~\bibnamefont {Lee}}, \bibinfo {author} {\bibfnamefont {R.}~\bibnamefont
  {Moore}}, \bibinfo {author} {\bibfnamefont {F.}~\bibnamefont {Schmitt}},
  \bibinfo {author} {\bibfnamefont {W.}~\bibnamefont {Li}}, \bibinfo {author}
  {\bibfnamefont {S.}~\bibnamefont {Riggs}}, \bibinfo {author} {\bibfnamefont
  {J.-H.}\ \bibnamefont {Chu}}, \bibinfo {author} {\bibfnamefont
  {B.}~\bibnamefont {Lv}}, \bibinfo {author} {\bibfnamefont {J.}~\bibnamefont
  {Hu}}, \bibinfo {author} {\bibfnamefont {M.}~\bibnamefont {Hashimoto}},
  \bibinfo {author} {\bibfnamefont {S.-K.}\ \bibnamefont {Mo}}, \bibinfo
  {author} {\bibfnamefont {Z.}~\bibnamefont {Hussain}}, \bibinfo {author}
  {\bibfnamefont {Z.}~\bibnamefont {Mao}}, \bibinfo {author} {\bibfnamefont
  {C.}~\bibnamefont {Chu}}, \bibinfo {author} {\bibfnamefont {I.}~\bibnamefont
  {Fisher}}, \bibinfo {author} {\bibfnamefont {Q.}~\bibnamefont {Si}}, \bibinfo
  {author} {\bibfnamefont {Z.-X.}\ \bibnamefont {Shen}},\ and\ \bibinfo
  {author} {\bibfnamefont {D.}~\bibnamefont {Lu}},\ }\bibfield  {title}
  {\bibinfo {title} {Observation of universal strong orbital-dependent
  correlation effects in iron chalcogenides},\ }\href
  {https://doi.org/10.1038/ncomms8777} {\bibfield  {journal} {\bibinfo
  {journal} {Nature Commun.}\ }\textbf {\bibinfo {volume} {6}},\ \bibinfo
  {pages} {7777} (\bibinfo {year} {2015})}\BibitemShut {NoStop}%
\bibitem [{\citenamefont {Yin}\ \emph {et~al.}(2011)\citenamefont {Yin},
  \citenamefont {Haule},\ and\ \citenamefont {Kotliar}}]{Yin:2011}%
  \BibitemOpen
  \bibfield  {author} {\bibinfo {author} {\bibfnamefont {Z.~P.}\ \bibnamefont
  {Yin}}, \bibinfo {author} {\bibfnamefont {K.}~\bibnamefont {Haule}},\ and\
  \bibinfo {author} {\bibfnamefont {G.}~\bibnamefont {Kotliar}},\ }\bibfield
  {title} {\bibinfo {title} {Kinetic frustration and the nature of the magnetic
  and paramagnetic states in iron pnictides and iron chalcogenides},\ }\href
  {https://doi.org/10.1038/nmat3120} {\bibfield  {journal} {\bibinfo  {journal}
  {Nature Mater.}\ }\textbf {\bibinfo {volume} {10}},\ \bibinfo {pages} {932}
  (\bibinfo {year} {2011})}\BibitemShut {NoStop}%
\bibitem [{\citenamefont {Stadler}\ \emph {et~al.}(2015)\citenamefont
  {Stadler}, \citenamefont {Yin}, \citenamefont {von Delft}, \citenamefont
  {Kotliar},\ and\ \citenamefont {Weichselbaum}}]{Stadler:2015}%
  \BibitemOpen
  \bibfield  {author} {\bibinfo {author} {\bibfnamefont {K.~M.}\ \bibnamefont
  {Stadler}}, \bibinfo {author} {\bibfnamefont {Z.~P.}\ \bibnamefont {Yin}},
  \bibinfo {author} {\bibfnamefont {J.}~\bibnamefont {von Delft}}, \bibinfo
  {author} {\bibfnamefont {G.}~\bibnamefont {Kotliar}},\ and\ \bibinfo {author}
  {\bibfnamefont {A.}~\bibnamefont {Weichselbaum}},\ }\bibfield  {title}
  {\bibinfo {title} {Dynamical mean-field theory plus numerical
  renormalization-group study of spin-orbital separation in a three-band hund
  metal},\ }\href {https://doi.org/10.1103/PhysRevLett.115.136401} {\bibfield
  {journal} {\bibinfo  {journal} {Phys. Rev. Lett.}\ }\textbf {\bibinfo
  {volume} {115}},\ \bibinfo {pages} {136401} (\bibinfo {year}
  {2015})}\BibitemShut {NoStop}%
\bibitem [{\citenamefont {Skornyakov}\ \emph {et~al.}(2017)\citenamefont
  {Skornyakov}, \citenamefont {Anisimov}, \citenamefont {Vollhardt},\ and\
  \citenamefont {Leonov}}]{Skornyakov:2017}%
  \BibitemOpen
  \bibfield  {author} {\bibinfo {author} {\bibfnamefont {S.~L.}\ \bibnamefont
  {Skornyakov}}, \bibinfo {author} {\bibfnamefont {V.~I.}\ \bibnamefont
  {Anisimov}}, \bibinfo {author} {\bibfnamefont {D.}~\bibnamefont
  {Vollhardt}},\ and\ \bibinfo {author} {\bibfnamefont {I.}~\bibnamefont
  {Leonov}},\ }\bibfield  {title} {\bibinfo {title} {Effect of electron
  correlations on the electronic structure and phase stability of fese upon
  lattice expansion},\ }\href {https://doi.org/10.1103/PhysRevB.96.035137}
  {\bibfield  {journal} {\bibinfo  {journal} {Phys. Rev. B}\ }\textbf {\bibinfo
  {volume} {96}},\ \bibinfo {pages} {035137} (\bibinfo {year}
  {2017})}\BibitemShut {NoStop}%
\bibitem [{\citenamefont {{Kretzschmar}}\ \emph {et~al.}(2016)\citenamefont
  {{Kretzschmar}}, \citenamefont {{B{\"o}hm}}, \citenamefont
  {{Karahasanovi{\'c}}}, \citenamefont {{Muschler}}, \citenamefont {{Baum}},
  \citenamefont {{Jost}}, \citenamefont {{Schmalian}}, \citenamefont
  {{Caprara}}, \citenamefont {{Grilli}}, \citenamefont {{Di Castro}},
  \citenamefont {{Analytis}}, \citenamefont {{Chu}}, \citenamefont {{Fisher}},\
  and\ \citenamefont {{Hackl}}}]{Kretzschmar:2016}%
  \BibitemOpen
  \bibfield  {author} {\bibinfo {author} {\bibfnamefont {F.}~\bibnamefont
  {{Kretzschmar}}}, \bibinfo {author} {\bibfnamefont {T.}~\bibnamefont
  {{B{\"o}hm}}}, \bibinfo {author} {\bibfnamefont {U.}~\bibnamefont
  {{Karahasanovi{\'c}}}}, \bibinfo {author} {\bibfnamefont {B.}~\bibnamefont
  {{Muschler}}}, \bibinfo {author} {\bibfnamefont {A.}~\bibnamefont {{Baum}}},
  \bibinfo {author} {\bibfnamefont {D.}~\bibnamefont {{Jost}}}, \bibinfo
  {author} {\bibfnamefont {J.}~\bibnamefont {{Schmalian}}}, \bibinfo {author}
  {\bibfnamefont {S.}~\bibnamefont {{Caprara}}}, \bibinfo {author}
  {\bibfnamefont {M.}~\bibnamefont {{Grilli}}}, \bibinfo {author}
  {\bibfnamefont {C.}~\bibnamefont {{Di Castro}}}, \bibinfo {author}
  {\bibfnamefont {J.~H.}\ \bibnamefont {{Analytis}}}, \bibinfo {author}
  {\bibfnamefont {J.-H.}\ \bibnamefont {{Chu}}}, \bibinfo {author}
  {\bibfnamefont {I.~R.}\ \bibnamefont {{Fisher}}},\ and\ \bibinfo {author}
  {\bibfnamefont {R.}~\bibnamefont {{Hackl}}},\ }\bibfield  {title} {\bibinfo
  {title} {{Critical spin fluctuations and the origin of nematic order in ${\rm
  Ba(Fe_{1-x}Co_x)_2As_2}$}},\ }\href {https://doi.org/10.1038/NPHYS3634}
  {\bibfield  {journal} {\bibinfo  {journal} {Nat. Physics}\ }\textbf {\bibinfo
  {volume} {12}},\ \bibinfo {pages} {560} (\bibinfo {year} {2016})}\BibitemShut
  {NoStop}%
\bibitem [{\citenamefont {Chibani}\ \emph
  {et~al.}(2021{\natexlab{a}})\citenamefont {Chibani}, \citenamefont {Farina},
  \citenamefont {Massat}, \citenamefont {Cazayous}, \citenamefont {Sacuto},
  \citenamefont {Urata}, \citenamefont {Tanabe}, \citenamefont {Tanigaki},
  \citenamefont {B{\"o}hmer}, \citenamefont {Canfield} \emph
  {et~al.}}]{chibani2021npj}%
  \BibitemOpen
  \bibfield  {author} {\bibinfo {author} {\bibfnamefont {S.}~\bibnamefont
  {Chibani}}, \bibinfo {author} {\bibfnamefont {D.}~\bibnamefont {Farina}},
  \bibinfo {author} {\bibfnamefont {P.}~\bibnamefont {Massat}}, \bibinfo
  {author} {\bibfnamefont {M.}~\bibnamefont {Cazayous}}, \bibinfo {author}
  {\bibfnamefont {A.}~\bibnamefont {Sacuto}}, \bibinfo {author} {\bibfnamefont
  {T.}~\bibnamefont {Urata}}, \bibinfo {author} {\bibfnamefont
  {Y.}~\bibnamefont {Tanabe}}, \bibinfo {author} {\bibfnamefont
  {K.}~\bibnamefont {Tanigaki}}, \bibinfo {author} {\bibfnamefont {A.~E.}\
  \bibnamefont {B{\"o}hmer}}, \bibinfo {author} {\bibfnamefont {P.~C.}\
  \bibnamefont {Canfield}}, \emph {et~al.},\ }\bibfield  {title} {\bibinfo
  {title} {Lattice-shifted nematic quantum critical point in fese 1- x s x},\
  }\href@noop {} {\bibfield  {journal} {\bibinfo  {journal} {npj Quantum
  Materials}\ }\textbf {\bibinfo {volume} {6}},\ \bibinfo {pages} {1} (\bibinfo
  {year} {2021}{\natexlab{a}})}\BibitemShut {NoStop}%
\bibitem [{\citenamefont {Karahasanovic}\ \emph {et~al.}(2015)\citenamefont
  {Karahasanovic}, \citenamefont {Kretzschmar}, \citenamefont {B\"ohm},
  \citenamefont {Hackl}, \citenamefont {Paul}, \citenamefont {Gallais},\ and\
  \citenamefont {Schmalian}}]{Karahasanovic2015_PRB92_075134}%
  \BibitemOpen
  \bibfield  {author} {\bibinfo {author} {\bibfnamefont {U.}~\bibnamefont
  {Karahasanovic}}, \bibinfo {author} {\bibfnamefont {F.}~\bibnamefont
  {Kretzschmar}}, \bibinfo {author} {\bibfnamefont {T.}~\bibnamefont {B\"ohm}},
  \bibinfo {author} {\bibfnamefont {R.}~\bibnamefont {Hackl}}, \bibinfo
  {author} {\bibfnamefont {I.}~\bibnamefont {Paul}}, \bibinfo {author}
  {\bibfnamefont {Y.}~\bibnamefont {Gallais}},\ and\ \bibinfo {author}
  {\bibfnamefont {J.}~\bibnamefont {Schmalian}},\ }\bibfield  {title} {\bibinfo
  {title} {{Manifestation of nematic degrees of freedom in the Raman response
  function of iron pnictides}},\ }\href
  {https://doi.org/10.1103/PhysRevB.92.075134} {\bibfield  {journal} {\bibinfo
  {journal} {Phys. Rev. B}\ }\textbf {\bibinfo {volume} {92}},\ \bibinfo
  {pages} {075134} (\bibinfo {year} {2015})}\BibitemShut {NoStop}%
\bibitem [{\citenamefont {Zhang}\ \emph {et~al.}(2021)\citenamefont {Zhang},
  \citenamefont {Wu}, \citenamefont {Kasahara}, \citenamefont {Shibauchi},
  \citenamefont {Matsuda},\ and\ \citenamefont {Blumberg}}]{ZhangWL:2021}%
  \BibitemOpen
  \bibfield  {author} {\bibinfo {author} {\bibfnamefont {W.}~\bibnamefont
  {Zhang}}, \bibinfo {author} {\bibfnamefont {S.}~\bibnamefont {Wu}}, \bibinfo
  {author} {\bibfnamefont {S.}~\bibnamefont {Kasahara}}, \bibinfo {author}
  {\bibfnamefont {T.}~\bibnamefont {Shibauchi}}, \bibinfo {author}
  {\bibfnamefont {Y.}~\bibnamefont {Matsuda}},\ and\ \bibinfo {author}
  {\bibfnamefont {G.}~\bibnamefont {Blumberg}},\ }\bibfield  {title} {\bibinfo
  {title} {{Quadrupolar charge dynamics in the nonmagnetic FeSe$_{1-x}$S$_x$
  superconductors}},\ }\bibfield  {journal} {\bibinfo  {journal} {Proc. Nat.
  Acad. Sci.}\ }\textbf {\bibinfo {volume} {118}},\ \href
  {https://doi.org/10.1073/pnas.2020585118} {10.1073/pnas.2020585118} (\bibinfo
  {year} {2021})\BibitemShut {NoStop}%
\bibitem [{\citenamefont {Chibani}\ \emph
  {et~al.}(2021{\natexlab{b}})\citenamefont {Chibani}, \citenamefont {Farina},
  \citenamefont {Massat}, \citenamefont {Cazayous}, \citenamefont {Sacuto},
  \citenamefont {Urata}, \citenamefont {Tanabe}, \citenamefont {Tanigaki},
  \citenamefont {B{\"o}hmer}, \citenamefont {Canfield}, \citenamefont {Merz},
  \citenamefont {Karlsson}, \citenamefont {Strobel}, \citenamefont
  {Toulemonde}, \citenamefont {Paul},\ and\ \citenamefont
  {Gallais}}]{Chibani2021}%
  \BibitemOpen
  \bibfield  {author} {\bibinfo {author} {\bibfnamefont {S.}~\bibnamefont
  {Chibani}}, \bibinfo {author} {\bibfnamefont {D.}~\bibnamefont {Farina}},
  \bibinfo {author} {\bibfnamefont {P.}~\bibnamefont {Massat}}, \bibinfo
  {author} {\bibfnamefont {M.}~\bibnamefont {Cazayous}}, \bibinfo {author}
  {\bibfnamefont {A.}~\bibnamefont {Sacuto}}, \bibinfo {author} {\bibfnamefont
  {T.}~\bibnamefont {Urata}}, \bibinfo {author} {\bibfnamefont
  {Y.}~\bibnamefont {Tanabe}}, \bibinfo {author} {\bibfnamefont
  {K.}~\bibnamefont {Tanigaki}}, \bibinfo {author} {\bibfnamefont {A.~E.}\
  \bibnamefont {B{\"o}hmer}}, \bibinfo {author} {\bibfnamefont {P.~C.}\
  \bibnamefont {Canfield}}, \bibinfo {author} {\bibfnamefont {M.}~\bibnamefont
  {Merz}}, \bibinfo {author} {\bibfnamefont {S.}~\bibnamefont {Karlsson}},
  \bibinfo {author} {\bibfnamefont {P.}~\bibnamefont {Strobel}}, \bibinfo
  {author} {\bibfnamefont {P.}~\bibnamefont {Toulemonde}}, \bibinfo {author}
  {\bibfnamefont {I.}~\bibnamefont {Paul}},\ and\ \bibinfo {author}
  {\bibfnamefont {Y.}~\bibnamefont {Gallais}},\ }\bibfield  {title} {\bibinfo
  {title} {{Lattice-shifted nematic quantum critical point in
  FeSe$_{1-x}$S$_x$}},\ }\href {https://doi.org/10.1038/s41535-021-00336-3}
  {\bibfield  {journal} {\bibinfo  {journal} {npj Quantum Materials}\ }\textbf
  {\bibinfo {volume} {6}},\ \bibinfo {pages} {37} (\bibinfo {year}
  {2021}{\natexlab{b}})}\BibitemShut {NoStop}%
\bibitem [{\citenamefont {Gonze}\ \emph {et~al.}(2020)\citenamefont {Gonze},
  \citenamefont {Amadon}, \citenamefont {Antonius}, \citenamefont {Arnardi},
  \citenamefont {Baguet}, \citenamefont {Beuken}, \citenamefont {Bieder},
  \citenamefont {Bottin}, \citenamefont {Bouchet}, \citenamefont {Bousquet}
  \emph {et~al.}}]{gonze2020}%
  \BibitemOpen
  \bibfield  {author} {\bibinfo {author} {\bibfnamefont {X.}~\bibnamefont
  {Gonze}}, \bibinfo {author} {\bibfnamefont {B.}~\bibnamefont {Amadon}},
  \bibinfo {author} {\bibfnamefont {G.}~\bibnamefont {Antonius}}, \bibinfo
  {author} {\bibfnamefont {F.}~\bibnamefont {Arnardi}}, \bibinfo {author}
  {\bibfnamefont {L.}~\bibnamefont {Baguet}}, \bibinfo {author} {\bibfnamefont
  {J.-M.}\ \bibnamefont {Beuken}}, \bibinfo {author} {\bibfnamefont
  {J.}~\bibnamefont {Bieder}}, \bibinfo {author} {\bibfnamefont
  {F.}~\bibnamefont {Bottin}}, \bibinfo {author} {\bibfnamefont
  {J.}~\bibnamefont {Bouchet}}, \bibinfo {author} {\bibfnamefont
  {E.}~\bibnamefont {Bousquet}}, \emph {et~al.},\ }\bibfield  {title} {\bibinfo
  {title} {The abinit project: Impact, environment and recent developments},\
  }\href@noop {} {\bibfield  {journal} {\bibinfo  {journal} {Computer Physics
  Communications}\ }\textbf {\bibinfo {volume} {248}},\ \bibinfo {pages}
  {107042} (\bibinfo {year} {2020})}\BibitemShut {NoStop}%
\bibitem [{\citenamefont {Perdew}\ \emph {et~al.}(2008)\citenamefont {Perdew},
  \citenamefont {Ruzsinszky}, \citenamefont {Csonka}, \citenamefont {Vydrov},
  \citenamefont {Scuseria}, \citenamefont {Constantin}, \citenamefont {Zhou},\
  and\ \citenamefont {Burke}}]{PhysRevLett.100.136406}%
  \BibitemOpen
  \bibfield  {author} {\bibinfo {author} {\bibfnamefont {J.~P.}\ \bibnamefont
  {Perdew}}, \bibinfo {author} {\bibfnamefont {A.}~\bibnamefont {Ruzsinszky}},
  \bibinfo {author} {\bibfnamefont {G.~I.}\ \bibnamefont {Csonka}}, \bibinfo
  {author} {\bibfnamefont {O.~A.}\ \bibnamefont {Vydrov}}, \bibinfo {author}
  {\bibfnamefont {G.~E.}\ \bibnamefont {Scuseria}}, \bibinfo {author}
  {\bibfnamefont {L.~A.}\ \bibnamefont {Constantin}}, \bibinfo {author}
  {\bibfnamefont {X.}~\bibnamefont {Zhou}},\ and\ \bibinfo {author}
  {\bibfnamefont {K.}~\bibnamefont {Burke}},\ }\bibfield  {title} {\bibinfo
  {title} {Restoring the density-gradient expansion for exchange in solids and
  surfaces},\ }\href {https://doi.org/10.1103/PhysRevLett.100.136406}
  {\bibfield  {journal} {\bibinfo  {journal} {Phys. Rev. Lett.}\ }\textbf
  {\bibinfo {volume} {100}},\ \bibinfo {pages} {136406} (\bibinfo {year}
  {2008})}\BibitemShut {NoStop}%
\bibitem [{\citenamefont {Hamann}(2013)}]{PhysRevB.88.085117}%
  \BibitemOpen
  \bibfield  {author} {\bibinfo {author} {\bibfnamefont {D.~R.}\ \bibnamefont
  {Hamann}},\ }\bibfield  {title} {\bibinfo {title} {Optimized norm-conserving
  vanderbilt pseudopotentials},\ }\href
  {https://doi.org/10.1103/PhysRevB.88.085117} {\bibfield  {journal} {\bibinfo
  {journal} {Phys. Rev. B}\ }\textbf {\bibinfo {volume} {88}},\ \bibinfo
  {pages} {085117} (\bibinfo {year} {2013})}\BibitemShut {NoStop}%
\bibitem [{\citenamefont {{van Setten}}\ \emph {et~al.}(2018)\citenamefont
  {{van Setten}}, \citenamefont {Giantomassi}, \citenamefont {Bousquet},
  \citenamefont {Verstraete}, \citenamefont {Hamann}, \citenamefont {Gonze},\
  and\ \citenamefont {Rignanese}}]{VANSETTEN201839}%
  \BibitemOpen
  \bibfield  {author} {\bibinfo {author} {\bibfnamefont {M.}~\bibnamefont {{van
  Setten}}}, \bibinfo {author} {\bibfnamefont {M.}~\bibnamefont {Giantomassi}},
  \bibinfo {author} {\bibfnamefont {E.}~\bibnamefont {Bousquet}}, \bibinfo
  {author} {\bibfnamefont {M.}~\bibnamefont {Verstraete}}, \bibinfo {author}
  {\bibfnamefont {D.}~\bibnamefont {Hamann}}, \bibinfo {author} {\bibfnamefont
  {X.}~\bibnamefont {Gonze}},\ and\ \bibinfo {author} {\bibfnamefont {G.-M.}\
  \bibnamefont {Rignanese}},\ }\bibfield  {title} {\bibinfo {title} {The
  pseudodojo: Training and grading a 85 element optimized norm-conserving
  pseudopotential table},\ }\href
  {https://doi.org/https://doi.org/10.1016/j.cpc.2018.01.012} {\bibfield
  {journal} {\bibinfo  {journal} {Computer Physics Communications}\ }\textbf
  {\bibinfo {volume} {226}},\ \bibinfo {pages} {39} (\bibinfo {year}
  {2018})}\BibitemShut {NoStop}%
\bibitem [{\citenamefont {Setyawan}\ and\ \citenamefont
  {Curtarolo}(2010)}]{SETYAWAN2010299}%
  \BibitemOpen
  \bibfield  {author} {\bibinfo {author} {\bibfnamefont {W.}~\bibnamefont
  {Setyawan}}\ and\ \bibinfo {author} {\bibfnamefont {S.}~\bibnamefont
  {Curtarolo}},\ }\bibfield  {title} {\bibinfo {title} {High-throughput
  electronic band structure calculations: Challenges and tools},\ }\href
  {https://doi.org/https://doi.org/10.1016/j.commatsci.2010.05.010} {\bibfield
  {journal} {\bibinfo  {journal} {Computational Materials Science}\ }\textbf
  {\bibinfo {volume} {49}},\ \bibinfo {pages} {299} (\bibinfo {year}
  {2010})}\BibitemShut {NoStop}%
\end{thebibliography}%

%\end{document}

%%%%%%%%%%%%%%%%%%%%%%%%%%%%%%%%%%%%%%%%%%%%%%%%

\clearpage
\begin{appendix}
\label{sec:appendix}

\setcounter{figure}{0}
\renewcommand\thefigure{A\arabic{figure}}

\setcounter{table}{0}
\renewcommand\thetable{A\Roman{table}}

\section{Phonon dispersion and density of states}
\label{Asec:PDOS}

We have performed density functional theory (DFT) calculations as implemented in the ABINIT package \cite{gonze2020}. We have used the Perdew-Burke-Ernzerhof functional tailored for solids (PBEsol) \cite{PhysRevLett.100.136406} and optimized norm-conserving pseudopotentials (ONCVPSP) \cite{PhysRevB.88.085117,VANSETTEN201839}, where Fe-3s$^2$3p$^6$3d$^6$4s$^2$, S-3s$^2$3p$^4$ and Se-3d$^{10}$4s$^2$4p$^4$ are treated as valence electrons. The energy cutoff for the planewave basis was set to 50 Ha. The lattice parameters and atomic positions used in the calculations were directly obtained from our X-ray diffraction measurements (performed at 300 K). Following previous first-principles studies on phonons in e.g.~FeS \cite{Baum:2018}, the crystal structures were not further relaxed, to achieve optimal characterization of the phonon frequencies. Here, both FeS and FeSe adopt the simple tetragonal space group \textit{P4/nmm} (No.~129), where Fe occupies \textit{2a} and S/Se position \textit{2c} Wyckoff position. The latter comprises an additional degree of freedom, namely the height of the chalcogen atoms S and Se with respect to the Fe plane, denoted as $z$. An overview of the lattice parameters that were used in the calculations is provided in Table~\ref{tab:latt_par}. 

\begin{table}[h]
\centering
\begin{tabular}{|c|c|c|c|}
\hline
Compound & $a$ (\AA) & $c$ (\AA) & $z$ (units of $c$) \\ \hline \hline
FeS & 3.6795 & 5.0321 & 0.2578 \\ \hline
FeSe & 3.7707 & 5.5202 & 0.2671 \\ \hline
\end{tabular}
\caption{Lattice parameters, obtained from X-ray diffraction measurements, used in the DFT and DFPT calculations.} 
\label{tab:latt_par}
\end{table}

Subsequently, the phonon dispersions were obtained from density functional perturbation theory (DFPT) calculations, also within ABINIT. Here, we have used a $15 \times 15 \times 9$ $\textit{k}$-point grid for the electron wave vectors and a $5 \times 5 \times 3$ $\textit{q}$-point grid for the phonon wave vectors. For the electronic occupation we employed Fermi-Dirac smearing with broadening factor $\sigma=0.01$ Ha.

\begin{figure*}[h]
\includegraphics[width=160mm]{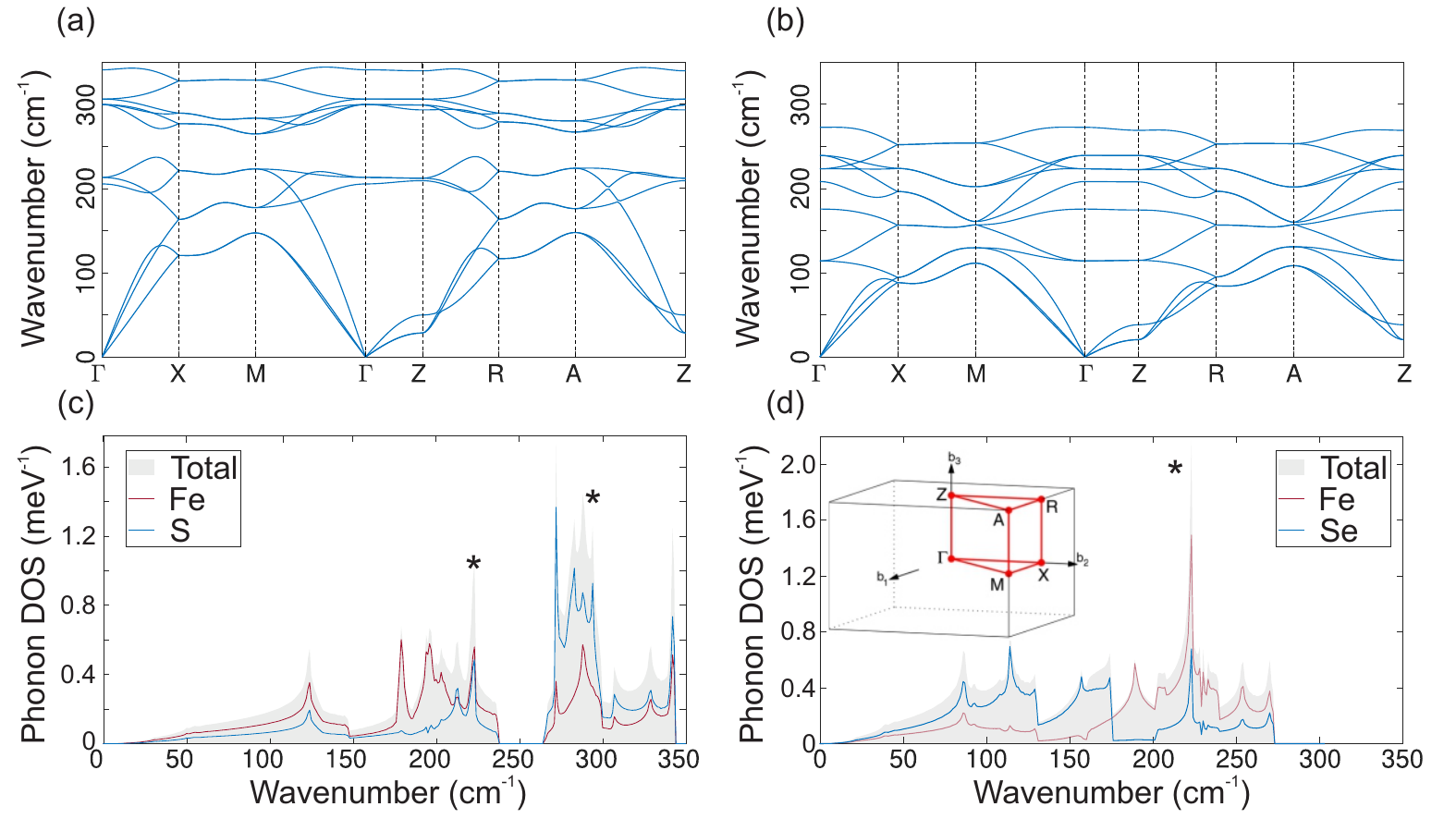}
\caption{Calculated phonon band structures of (a) FeS and (b) FeSe. Phonon DOS of (c) FeS and (d) FeSe, including partial contributions from Fe and S/Se. Brillouin zone of both structures depicted in black is shown in Inset, with the irreducible Brillouin zone, along which the band structures are plotted, in red \cite{SETYAWAN2010299}. The energies of the extra peaks in Fig. \ref{Afig:more} are also indicated here by asterix. }
\label{fig_S1}
\end{figure*}

The results of these calculations are shown in Fig~\ref{fig_S1}. FeS is found to have phonon frequencies stretching up to 344 \wn [Fig\ref{fig_S1} (a)], which is significantly higher than the maximum phonon value of 273 \wn obtained for FeSe [Fig \ref{fig_S1}(b)], owing to the higher atomic mass of Se compared to S. The atom-resolved phonon densities of states (DOS) of both compounds reveal a mixture of iron and chalcogen contribution throughout the entire phonon spectrum [Fig. \ref{fig_S1}(c) and (d)]. Interestingly, there is a change of dominant phonon character, with the lower modes dominated by Fe in FeS, while the lower modes have predominant Se character in FeSe. This reversal can be understood from the fact that the atomic number of Fe ($Z=26$) lies in between those of S ($Z=16$) and Se ($Z=34$). These differences in atomic masses lead moreover to a small energy gap between Fe- and S-dominated modes in FeS (between 238 and 265 \wn), which is entirely absent in FeSe.

%%%%%%%%%%%%%%%%%%%%%%%%%%%%%%%%%%%%%%%%%%%%%%%%%%%%%%%%%%
%\begin{figure*}
%  \centering
%  \includegraphics[width=140mm]{./FigureA1}
%  \caption[]{(Color online) Phonon dispersion along the main symmetry lines (a) and (b) and phonon density of states (c) and (d) for FeSe and FeS. The gap between acoustical and optical modes is indicated and is fully developed only for FeS. The energies of the extra peaks in Fig. \ref{Afig:more} are also indicated here.}
%  \label{Afig:dispersion}
%\end{figure*}
%%%%%%%%%%%%%%%%%%%%%%%%%%%%%%%%%%%%%%%%%%%%%%%%%%%%%

%%%%%%%%%%%%%%%%%%%%%%%%%%%%%%%%%%%%%%%%%%%%%%%%%%%%%%%%%%
%\begin{figure*}
%  \centering
%  \includegraphics[width=140mm]{./FigureA2}
%  \caption[]{(Color online) PDOS for FeSe and FeS. The energies of the extra peaks in Fig. \ref{Afig:more} are also indicated here.}
%  \label{Afig:PDOS}
%\end{figure*}
%%%%%%%%%%%%%%%%%%%%%%%%%%%%%%%%%%%%%%%%%%%%%%%%%%%%%

\section{FeSe$_{0.52}$S$_{0.48}$ and FeS}
\label{Asec:more doping}
\begin{figure}
  \centering
  \includegraphics[width=85mm]{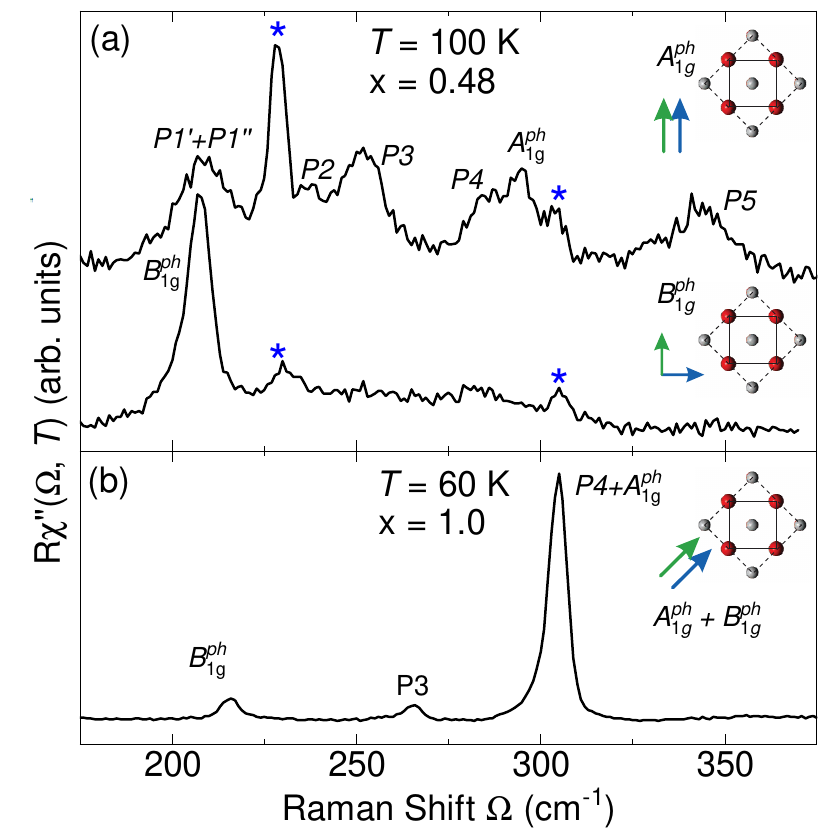}
  \caption[]{(Color online) \FESES for $x=0.48$ and $x=1$. (a) \Algph ($xx$) and \Blgph ($xy$) spectra for FeSe$_{0.52}$S$_{0.48}$. In addition to the phonons and the structures observed at the other doping levels there are two relatively sharp lines (marked by asterisks) which we associate with the PDOS. They may also arise from a nearly ordered super-structure close to 50\% doping.}
  \label{Afig:more}
\end{figure}

For convenience we show here additional doping levels in Fig \ref{Afig:more}. The spectrum for $x=1$ in panel (b) was already published elsewhere \cite{Baum:2018}. Note that for $x^\prime x^\prime$ both \Algph and \Blgph are projected and that the labels for the symmetry-forbidden peaks P3 and P4 are different from those in the earlier paper. $x=0.48$ [Fig \ref{Afig:more} (a)] is in the middle between FeSe and FeS, and one can therefore expect the strongest contribution from defect-induced scattering. This interpretation is supported by the presence of structures in both configurations. All peaks resolved at $x=0.69$ in $xx$ configuration are also observed here. In addition there are two lines marked by asterisks which appear only at $x=0.48$. Since they appear also for $x^\prime x^\prime$ we interpret them in terms of contributions from the PDOS as shown in Fig. \ref{fig_S1} where the respective energies correspond to a high DOS of either FeSe or FeS. Structure P5 may be related to the high-energy part of FeS. 
%%%%%%%%%%%%%%%%%%%%%%%%%%%%%%%%%%%%%%%%%%%%%%%%%%%%%%%%%
%%%%%%%%%%%%%%%%%%%%%%%%%%%%%%%%%%%%%%%%%%%%%%%%%%%%

\end{appendix}

\end{document}